\documentclass{aastex631}

\usepackage{amsmath}
\usepackage{empheq}
\usepackage{multirow}

\begin{document}

\title{Kerr-scalaron metric and astronomical consequences near the Galactic Center black hole}

\correspondingauthor{Debojit Paul}
\email{debojitpaul645@gmail.com, debojit@gauhati.ac.in}

\author[0000-0003-4301-3496]{Debojit Paul}
\affiliation{Department of Physics, Gauhati University \\
Jalukbari, Guwahati-781014 \\
Assam, India}

\author[0009-0005-5021-9996]{Pranjali Bhattacharjee}
\affiliation{Department of Physics, Gauhati University \\
Jalukbari, Guwahati-781014 \\
Assam, India}

\author[0000-0002-2880-4284]{Sanjeev Kalita}
\affiliation{Department of Physics, Gauhati University \\
Jalukbari, Guwahati-781014 \\
Assam, India}

%\collaboration{20}{(AAS Journals Data Editors)}



\begin{abstract}

Astronomical tests of spacetime metric and gravitation theory near the Galactic Center (GC) black hole, Sgr A* have gained momentum with the observations of compact stellar orbits near the black hole and measurement of the black hole shadow. Deviation from the Kerr metric is a potential signature of modified gravity theory. In this work, we use Newman-Janis algorithm to construct an axially symmetric and asymptotically flat metric in f(R) scalaron gravity theory. We call it as Kerr-scalaron metric. For studying astronomical consequences of the new metric we use the compact stellar orbits and the black hole shadow. We use the observed size of the emission ring of the GC black hole shadow for estimating deviation of the new metric from general relativity. It has been found that scalarons with mass within $10^{-17}$ eV - $10^{-16}$ eV are compatible with the observed emission ring size for black hole spin $\chi=0.9$. Schwarzschild limit of the pericenter shift is estimated for compact stellar orbits near the black hole. General relativistic  pericenter shift in wider orbits including S-stars such as S4716 and S2 has been reproduced with these scalarons. The parameter $f_{SP}$ measuring deviation from Schwarzschild pericenter shift has been found as $f_{SP}=1.00-1.04$ within stellar orbits having semi-major axes $45$ au - $100$ au. Scalarons have the capability to dominate Schwarzschild precession for orbits much below $45$ au. Lense-Thirring (LT) precession with the new metric is estimated for the compact orbits. The massive scalarons produce LT precession with magnitude ($12.25-24.5$) $\mu$as/yr in the orbit of S2. The LT precession time scale is within $0.1$\% of the age of the S-stars.

\end{abstract}



\section{Introduction} \label{sec:intro}

The no-hair theorem states that black holes in general relativity (GR) are uniquely described by three parameters – mass, spin and electric charge \citep{PhysRev.164.1776,PhysRevLett.26.331,hawking1972black} encapsulated in the Kerr-Newman metric \citep{PhysRevLett.11.237,10.1063/1.1704351}. Astrophysical black holes being deprived of net electric charge are described by the stationary, axisymmetric and asymptotically flat Kerr metric \citep{PhysRevLett.11.237} which contains only two parameters – mass and spin of the black hole. Therefore, presence of extra degree of freedom in addition to mass and spin is a signature of potential deviation from GR. Astronomical tests of departure from the Kerr metric are crucial for constraining alternative gravitation theories.

\par Several astronomical tests have been suggested in literature to examine whether black holes are described by the Kerr metric. These include gravitational waves (see \cite{Hughes:2010xf} for review;\citet{KONOPLYA2016350,berti2019topical}), electromagnetic emission due to accretion flows near black holes \citep{Johannsen_2010a}, pulsar-black hole binaries \citep{Wex_1999,10.1093/mnras/stac2337,10.1093/mnras/stad2125} and ephemerides of stars encircling the Galactic Center (GC) black hole, Sgr A* \citep{Will_2008}. \citet{Zhang_2015} discussed possibility of constraining spin and mass of the GC black hole through orbital dynamics  of six example S-stars in the nuclear star cluster and propagation of photons from a star orbiting the hole. They reported that astrometric and spectroscopic precisions of orders $10\ \mu$as and $1$ km/sec respectively to be achieved by the GRAVITY interferometer on the Very Large Telescope (VLT), upcoming Thirty Meter Telescope (TMT) and other Extremely Large Telescopes (ELTs) will be able to constrain spin of the black hole through less than two decades of tracking of the S2 star. This is expected to enable astronomers to test the Kerr metric hypothesis and the no-hair theorem. The GRAVITY interferometer combined with the SINFONI spectrograph at VLT is expected to be capable enough for detecting several general relativistic effects(including the Lense-Thirring effect \citep{lense1918influence} near the GC black hole) in the orbit of S2 within a very short observation window \citep{grould_2017}.

\par Tests of modified gravity theory and spacetime structure near the GC black hole and the M87* black hole (the supermassive black hole with mass of around $10^9 M_\odot$ residing at the center of the massive elliptical galaxy, M87 and being nearest to the GC black hole) constitute an interesting avenue for gravitational physics, cosmology and observational astronomy. These tests are composed of monitoring short period stars (period of the order of a decade) in the nuclear star cluster near the GC black hole and measuring the angular size and shape of the silhouette of the two black holes \citep{PhysRevLett.118.211101,gravity2020,PhysRevD.104.L101502,Akiyama_2019,Akiyama_VI}. The bright emission ring and the angular size of the GC black hole has been used to constrain a wide class of gravitational theories \citep{Vagnozzi_2023}.

\par There are two approaches to test modified gravity theories. In one approach one assumes a specific theory of gravity and produces constraints on deviation from GR by interpreting observations within the regime of that theory. Conventional strong field tests of gravity rely on this approach \citep{Johannsen_2016}. Einstein – Maxwell – Dilaton – Axion (EMDA) gravity theory has been recently constrained with the orbit of S2 \citep{Fernandez:2023kro}. It has been possible through the spherically symmetric black hole metric in EMDA gravity. These are interesting alternatives to GR appearing from heterotic string theory which also predict peculiar observational features in direct imaging of black holes. Currently publicly available data on the orbit of S2 has been used put constraint on the theory.

Alternatively one considers generic deviation from GR so as to encompass a wide class of modified gravity theories. Weak field limit of these theories is parameterized by certain parameters which are then constrained by observations in such a way that all those modified theories are constrained in one go. This is the well known Parameterized Post Newtonian (PPN) approach. Standard solar system tests of scalar –tensor gravity theories \citep{PhysRevD.93.044013,faraoni2020solar,gonzalez2020constraints} and constraints put by Schwarzschild precession of the S2 star \citep{gravity2020} encircling the GC black hole in a 16 year orbit are based on this category.  In strong field environment one cannot apply PPN tests.  Fortunately the Kerr metric is not unique to GR. There are alternative theories of gravity permitting Kerr like vacuum solutions. Examples include Randall-Sundrum type braneworld gravity \citep{PhysRevD.71.104027}, Einstein-dilaton-Gauss-Bonnet gravity \citep{mignemi1993dilaton}, Chern-Simons gravity \citep{PhysRevD.84.087501}, Horava-Lifhsitz gravity \citep{PhysRevD.83.124043} and Horndeski gravity \citep{PhysRevD.92.104049}.  Therefore, one may consider deviation from general relativistic Kerr metric and test the underlying theory. \citet{PhysRevD.83.124015} (JP) constructed a metric containing parametric deviation from the Kerr metric which is eligible to perform strong field tests of gravity. The deviation parameters of the JP metric can be mapped to any theory of gravity. Recent measurement of the angular size and shape of the GC black hole shadow has been used to constrain the deviation parameters of the JP metric \citep{Akiyama_VI}.   

Direct constraints on some metrics have appeared after imaging of the M87* and Sgr A* black holes. Tidal charge present in higher dimensional gravity theory \citep{dadhich2000black,universe8030141} and de Sitter background (space with a positive cosmological constant) affect the Schwarzschild and Kerr metrics (Schwarzschild metric can be converted to Kerr metric through the Newman – Janis algorithm (NJA) (see \citet{1965JMP.....6..915N}). Alteration of the metric causes shift of the angular size of the photon radius and hence the size of the black hole shadow. Constraints on tidal charge and the cosmological constant have been realized through the angular size of the shadows of M87* and Sgr A* black holes \citep{neves2020constraining,universe8030141,kalita2023constraining}. The M87* black hole shadow has also been used to generate constraints on ultralight scalar particles which are known as scalar hair of black holes \citep{universe5120220}.

f(R) gravity theory is based on perturbations to the Ricci scalar (R) in the Einstein-Hilbert action. It contains a scalar mode of gravity known as scalaron . These theories have been extensively studied in the context of dark energy \citep{capozziello2002curvature,capozziello2003curvature,PhysRevD.70.043528,PhysRevD.68.123512,PhysRevD.76.064004} and dark matter \citep{10.1111/j.1365-2966.2007.11401.x,PhysRevD.90.044052}. Several observational tests of the f(R) theories in cosmology have been performed by using cosmological probes such as expansion history and structure formation \citep{gu2011cosmological}, quasar samples \citep{Xu_2018}, cosmic void \citep{PhysRevD.104.023512}, type Ia supernovae \citep{hough2020viability} and galaxy clustering ratio \citep{PhysRevD.91.103503}. After the remarkable detection of Schwarzschild precession of S2 by the \citep{gravity2020} some parameters of the f(R) gravity theories were constrained by the orbital data of S2 \citep{PhysRevD.104.L101502}. By analyzing the orbit of S2, \citet{borka2016probing} reported constraints on power law f(R) gravity with $f(R)\sim R^m$ . Competition between f(R) gravity and GR in prediction of pericenter shift of compact stellar orbits below the orbit of S2 and measurability through GRAVITY beam combiner on VLT, upcoming TMT and other ELTs have been extensively studied by \citet{Kalita_2020,Kalita_2021,doi:10.1142/S0218271823500219}.

Black holes in presence of scalar field have been extensively studied in the context of dark energy and no-hair theorem \citep{Kiselev:2002dx,PhysRevD.70.084035,Fernando:2012ue,PhysRevLett.112.251102}. Scalar fields are eligible to form bound state near a black hole if the Compton wavelength of the scalar particles exceeds the horizon length of the black hole \citep{Brito:2015oca,PhysRevLett.119.041101}. The possibility of detection of these scalar clouds through gravitational waves emitted from inspiralling binary black holes have also been discussed \citep{PhysRevD.99.044001,PhysRevD.99.104039}. Scalar field accretion has been studied for taking into account growth of the black holes and their connections with the background cosmology \citep{PhysRevLett.83.2699,2005JETP..100..528B,PhysRevD.80.104018}. Gravitation collapse of scalar field leading to singularity has also been studied \citep{Carneiro:2018url}. The prospect of constraining these scalar field configurations through the orbit of S2 have also been widely explored \citep{PhysRevD.108.L101303,DM2023,2019MNRAS.489.4606G,10.1093/mnras/stad1939}. 

\par In this work we construct an asymptotically flat Kerr metric in f(R) gravity theory containing the scalaron degree of freedom and study its astronomical consequences near the GC black holes. It is to be noted that in weak field limit these theories cause a Yukawa correction ($e^{-M_\psi r}/r$, with $M_\psi$ being the scalaron mass) to the Kepler potential \citep{2018ApJ...855...70K}. Scalarons are natural outcomes of curvature corrected quantum vacuum fluctuations near black holes. The UV and IR cut off of these fluctuations are determined by the scale of the Schwarzschild length and thermal energy density due to Hawking effect \citep{Kalita_2020}. The Kerr metric in f(R) theories has been constructed from the seed Schwarzschild metric in the theory (see \citet{2018ApJ...855...70K}) by employing the NJA. Asymptotic flatness of the metric has been ensured with a reciprocity relation between scalaron mass ($M_\psi$) and black hole mass (M). We call it as the Kerr-scalaron metric. Effective potential, orbit equation for test particles near the GC black hole and deviation of Schwarzschild pericenter shift from GR predictions for compact stellar orbits near the black hole have been studied. Size of the GC black hole shadow has been estimated with the Kerr-scalaron metric and scalaron mass has been constrained with the help of observed EHT bound on the angular size of the emission ring. Additionally, the Lense-Thirring precession of the stellar orbits is also studied in the new metric.

The paper is organized as follows. In section \ref{sec2} we develope the Kerr-scalaron metric with the NJA and present its effects such as effective potential, orbit equation, pericenter shift and black hole shadow size. In section \ref{sec3}, the astronomical prospects of the Kerr-scalaron metric near the GC black hole are discussed. Section \ref{results} presents results and discussions. Section \ref{summary} concludes the main results.

\section{Formalism of Kerr-scalaron Metric} \label{sec2}

The $f(R)$ theories begin with a geometrical modification to Einstein-Hilbert action such that the action becomes non-linear in $R$. The action has the form \citep{amendola2010dark}

\begin{equation}\label{eq1}
S=\int \frac{d^4x}{16\pi G}\sqrt{-g}f(R)+S_m(\phi_m,g_{\mu\nu}) ,
\end{equation} 

where, $f(R)\propto R^n$ ($n=1$ corresponds to GR and $n\neq 1$ corresponds to modified gravity) and $S_m$ is the matter part with matter fields $\phi_m$ coupled to spacetime metric $g_{\mu\nu}$. Starting from the field equation obtained by varying the action (\ref{eq1}) with respect to $g_{\mu\nu}$, a spherically symmetric metric has been obtained as \citep{2018ApJ...855...70K}

\begin{equation}\label{eq2}
ds^2=\left[1-\frac{2m}{r}\left(1+\frac{1}{3}e^{-M_\psi r}\right)\right]c^2dt^2-\left[1-\frac{2m}{r}\left(1+\frac{1}{3}e^{-M_\psi r}\right)\right]^{-1}dr^2-r^2d\Omega ,
\end{equation}

where, $m=GM/c^2\psi_o$ with $\psi_o$ being the background scalar field, $\psi_o=f'(R_o)$ known as scalaron ($R_o$ being the background curvature). Around a general relativistic background ($f(R) \approx R$) we consider the background field amplitude $\psi_o$ to be close to unity. $M_\psi$ is the mass of scalarons which acts like additional degree of freedom in f(R) gravity. We call metric (\ref{eq2}) as Schwarzschild-scalaron(SchS) metric. It is evident from equation (\ref{eq2}) that for infinitely large scalaron mass ($M_\psi\rightarrow \infty$), the metric reduces to Schwarzschild metric. The asymptotic flatness also follows. Through consideration of ultraviolet(UV) and infrared(IR) cut-off scales of curvature corrected quantum vaccum fluctuations in black hole spacetime, \citet{Kalita_2020} related the scalaron mass with horizon lenght of a black hole as $M_\psi \propto 1/m$ (see details in section \ref{sec2.2}). Therefore for $r>>m$ the SchS metric reduces to asymptotically flat form. In this section, we obtain the metric for an astrophysical black hole in presence of scalar fields. It is the axially symmetric and asymptotically flat metric generated from the spherically symmetric one (equation \ref{eq2}) through the method proposed by \citet{1965JMP.....6..915N} (NJA). In section \ref{sec2.1} we present the NJA formalism by following \citet{2010CQGra..27p5008C}.

\subsection{The NJA formalism} \label{sec2.1}
Any static and spherically symmetric metric is expressed as

\begin{equation}\label{eq3}
ds^2=e^{2\phi(r)}c^2dt^2-e^{2\lambda(r)}dr^2-r^2d\Omega ,
\end{equation}

where, $g_{tt}=e^{2\phi(r)}$ and $g_{rr}=-e^{2\lambda(r)}$. The metric is first transformed into Eddington-Finkelstein coordinates ($u,r,\theta,\phi$). This is acheived by transforming the time coordinate as $cdt=du\pm e^{\lambda(r)-\phi(r)}dr$. Under such transformation the metric takes the form

\begin{equation}\label{eq4}
ds^2=e^{2\phi(r)}du^2\pm2e^{\lambda(r)+\phi(r)}dudr-r^2d\Omega .
\end{equation}

For this line element, the contravariant metric tensor components are written as

\begin{equation}\label{eq5}
g^{\mu\nu}=\left(
\begin{array}{cccc}
0 & \pm e^{-\lambda(r)-\phi(r)} & 0 & 0 \\
\pm e^{-\lambda(r)-\phi(r)} & -e^{-2\lambda(r)} & 0 & 0 \\
0 & 0 & -\frac{1}{r^2} & 0 \\
0 & 0 & 0 & -\frac{1}{r^2 \sin^2\theta}
\end{array}
\right).
\end{equation}

The metric tensor $g^{\mu\nu}$ can be expressed in terms of null tetrads as

\begin{equation}\label{eq6}
g^{\mu\nu}=l^\mu n^\nu + l^\nu n^\mu - m^\mu \bar{m}^\nu-m^\nu \bar{m}^\mu ,
\end{equation}

where $l^\mu,n^\mu,m^\mu$ and $\bar{m}^\mu$ (the bar represents complex conjugate of $m^\mu$) are the null tetrads. These tetrads satisfy the conditions
 
\begin{equation}\label{eq7}
l_\mu l^\mu=m_\mu m^\mu = n_\mu n^\mu , \ l_\mu n^\mu=-m_\mu \bar{m}^\mu=1 ,\  l_\mu m^\mu=n_\mu m^\mu =0.
\end{equation}
 
Hence, the metric (\ref{eq5}) can be represented in terms of the null tetrads as

\begin{equation}\label{eq8}
	\begin{aligned}
		& l^\mu=\delta_1^\mu \\
		& n^\mu=e^{-\lambda(r)-\phi(r)}\delta_0^\mu-\frac{1}{2} e^{-2\lambda(r)}\delta_1^\mu \\
		& m^\mu=\frac{1}{\sqrt{2}r}\left(\delta_2^\mu+\frac{i}{\sin\theta}\delta_3^\mu\right) \\
		& \bar{m}^\mu=\frac{1}{\sqrt{2}r}\left(\delta_2^\mu-\frac{i}{\sin\theta}\delta_3^\mu\right).
	\end{aligned}
\end{equation}

The radial coordinate $r$ in $x^\mu=(u,r,\theta,\phi)$ is transformed to take complex values so that the tetrads become 

\begin{equation}\label{eq9}
	\begin{aligned}
		& l^\mu=\delta_1^\mu \\
		& n^\mu=e^{-\lambda(r,\bar{r})-\phi(r,\bar{r})}\delta_0^\mu-\frac{1}{2} e^{-2\lambda(r,\bar{r})}\delta_1^\mu \\
		& m^\mu=\frac{1}{\sqrt{2}\bar{r}}\left(\delta_2^\mu+\frac{i}{\sin\theta}\delta_3^\mu\right) \\
		& \bar{m}^\mu=\frac{1}{\sqrt{2}r}\left(\delta_2^\mu-\frac{i}{\sin\theta}\delta_3^\mu\right).
	\end{aligned}
\end{equation}

For $r=\bar{r}$ ($\bar{r}$ is the complex conjugate of $r$), tetrads in (\ref{eq9}) reduce to those in (\ref{eq8}) giving us the original metric in equation (\ref{eq5}). Under a complex coordinate transfomation of the form $x^\mu \rightarrow \tilde{x}^\mu=x^\mu+iy^\mu(x^\sigma)$, a new metric is obtained. Here, $y^\mu(x^\sigma)$ are the analytic functions of the real coordinates, $x^\sigma$. Simultaneously the null tetrads $Z_i^\mu=(l^\mu,n^\mu,m^\mu,\bar{m}^\mu)$ (here, $i=1,2,3,4$) transform as $Z_i^\mu\rightarrow \tilde{Z}_i^\mu(\tilde{x}^\sigma,\bar{\tilde{x}}^\sigma)=Z_i^\rho \frac{\delta \tilde{x}^\mu}{\delta x^\rho}$. It is essential to recover the old tetrads and the metric under the condition, $\tilde{x}^\sigma=\bar{\tilde{x}}^\sigma$. In order to achieve this goal, the following choice is considered

\begin{equation}\label{eq10}
\tilde{x}^\mu=x^\mu+ia(\delta_1^\mu-\delta_0^\mu) \cos\theta.
\end{equation}

As a result, the transformation of $x^\mu(u,r,\theta,\phi)\rightarrow \tilde{x}^\mu(\tilde{u},\tilde{r},\tilde{\theta},\tilde{\phi})$ has the form

\begin{equation}\label{eq11}
	\begin{aligned}
	& \tilde{u}=u+ia \cos \theta \\
	& \tilde{r}=r-ia \cos\theta \\
	& \tilde{\theta}=\theta \\
	& \tilde{\phi}=\phi.
	\end{aligned}
\end{equation}

where $a$ is constant parameter. For $\tilde{r}=\bar{\tilde{r}}$, the null tetrads in equation (\ref{eq9}) transform as

\begin{equation}\label{eq12}
	\begin{aligned}
		& \tilde{l}^\mu=\delta_1^\mu \\
		& \tilde{n}^\mu=e^{-\lambda(\tilde{r},\theta)-\phi(\tilde{r},\theta)}\delta_0^\mu-\frac{1}{2} e^{-2\lambda(\tilde{r},\theta)}\delta_1^\mu \\
		& \tilde{m}^\mu=\frac{1}{\sqrt{2}(\tilde{r}-ia\cos\theta)}\left[ia(\delta_0^\mu-\delta_1^\mu)\sin\theta+\delta_2^\mu+\frac{i}{\sin\theta}\delta_3^\mu\right] \\
		& \bar{\tilde{m}}^\mu=\frac{1}{\sqrt{2}(\tilde{r}+ia\cos\theta)}\left[-ia(\delta_0^\mu-\delta_1^\mu)\sin\theta+\delta_2^\mu-\frac{i}{\sin\theta}\delta_3^\mu\right].
	\end{aligned}
\end{equation}

Under the transformation equations (\ref{eq11}) and using the transformed null tetrads of equation (\ref{eq12}) in equation (\ref{eq6}), a new metric is obtained in the ($\tilde{u},\tilde{r},\theta,\phi$) coordinate which has the form

\begin{equation}\label{eq13}
\tilde{g}^{\mu\nu}=\left(
\begin{array}{cccc}
-\frac{a^2\sin^2\theta}{\rho^2} & e^{-\lambda(\tilde{r},\theta)-\phi(\tilde{r},\theta)}+\frac{a^2\sin^2\theta}{\rho^2} & 0 & -\frac{a}{\rho^2}\\
. & -e^{-2\lambda(\tilde{r},\theta)}-\frac{a^2\sin^2\theta}{\rho^2} & 0 & \frac{a}{\rho^2}\\
. & . & -\frac{1}{\rho^2} & 0 \\
. & . & . & -\frac{1}{\rho^2\sin^2\theta}
\end{array}\right),
\end{equation}

where $\rho^2=\tilde{r}^2+a^2\cos^2\theta$. Due to the symmetric nature of the metric, dots are used to represent $g^{\mu\nu}=g^{\nu\mu}$. Now, the covariant form of the metric (\ref{eq13}) is expressed as

\begin{equation}\label{eq14}
\tilde{g}_{\mu\nu}=\left(
\begin{array}{cccc}
e^{2\phi(\tilde{r},\theta)} & e^{\lambda(\tilde{r},\theta)+\phi(\tilde{r},\theta)} & 0 & ae^{\phi(\tilde{r},\theta)}[e^{\lambda(\tilde{r},\theta)}-e^{\phi(\tilde{r},\theta)}]\sin^2\theta \\
. & 0 & 0 & -ae^{\phi(\tilde{r},\theta)+\lambda(\tilde{r},\theta)}\sin^2\theta \\
. & . & -\rho^2 & 0 \\
. & . & . & -[\rho^2+a^2\sin^2\theta e^{\phi(\tilde{r},\theta)}(2e^{\lambda(\tilde{r},\theta)-e^{\phi(\tilde{r},\theta)}})]\sin^2\theta
\end{array}\right).
\end{equation}

Equation (\ref{eq14}) is the general form of the new metric after application of NJA to a sphreically symmetric 'seed' metric. This metric is further simplified by certain gauge transformations such that $g_{\phi t}$ is the only surviving off-diagonal term in the metric. This transformation is carried out so as to make the metric comparable to standard form of Kerr metric in Boyer-Lindquist coodinate system \citep{hobson2006general}. In order to achieve this, the coordinates $\tilde{u}$ and $\tilde{\phi}$ are redefined as

\begin{equation}\label{eq15}
d\tilde{u}=dt + g(\tilde{r})d\tilde{r}\hspace{0.5cm}  \text{and}\hspace{0.5cm}  d\tilde{\phi}=d\phi +h(\tilde{r})d\tilde{r}.
\end{equation}

where, 

\begin{equation}\label{eq16}
\begin{aligned}
\begin{cases}
& g(\tilde{r})=-\frac{e^{\lambda(\tilde{r},\theta)}(\rho^2+a^2 \sin^2\theta e^{\lambda(\tilde{r},\theta)+\phi(\tilde{r},\theta)})}{e^{\phi(\tilde{r},\theta)}(\rho^2+a^2\sin^2\theta e^{2\lambda(\tilde{r},\theta)})} \\
& h(\tilde{r})=-\frac{a e^{2\lambda(\tilde{r},\theta)}}{\rho^2+a^2\sin^2\theta e^{2\lambda(\tilde{r},\theta)}}
\end{cases}
\end{aligned}
\end{equation}

Applying the coordinate transformations (\ref{eq15}), the metric (\ref{eq14}) in the ($t,\tilde{r},\theta,\phi$) coordinate system takes the form

\begin{equation}\label{eq17}
g_{\mu\nu}=\left(
\begin{array}{cccc}
e^{2\phi(\tilde{r},\theta)} & 0 & 0 & a e^{\phi(\tilde{r},\theta)}[e^{\lambda(\tilde{r},\theta)}-e^{\phi(\tilde{r},\theta)}]\sin^2\theta \\
. & -\frac{\rho^2}{\rho^2 e^{-2\lambda(\tilde{r},\theta)}+a^2\sin^2\theta} & 0 & 0 \\
. & . & -\rho^2 & 0 \\
. & . & . & -[\rho^2+a^2\sin^2\theta e^{\phi(\tilde{r},\theta)}(2 e^{\lambda(\tilde{r},\theta)}-e^{\phi(\tilde{r},\theta)})]\sin^2\theta
\end{array}\right).
\end{equation}

Equation (\ref{eq17}) corresponds to complete family of axially symmetric metric obtained after applying the NJA. It is to be noted that the transformations defined in equations (\ref{eq15}) and (\ref{eq16}) are valid under the condition, $\rho^2+a^2\sin^2\theta e^{2\lambda(\tilde{r},\theta)}\neq 0$ where, $e^{2\lambda(\tilde{r},\theta)} \ge 0$. This technique is applied in section \ref{sec2.2} to generate an axially symmetric and asymptotically flat metric from the spherically symmetric metric (\ref{eq2}).

\subsection{The Kerr-scalaron metric}\label{sec2.2}

Following metric (\ref{eq5}), the contravariant metric tensor components for the metric (\ref{eq2}) in Eddington-Finkelstein coordinates ($u,r,\theta,\phi$) are expressed as

\begin{equation}\label{eq18}
g^{\mu\nu}=\left(\begin{array}{cccc}
0 & 1 & 0 & 0 \\
1 & -\left(1-\frac{2m}{r}\alpha(r)\right) & 0 & 0 \\
0 & 0 & -1/r^2 & 0\\
0 & 0 & 0 & -1/(r^2 \sin^2\theta) .
		\end{array}\right).
\end{equation} 

where, $\alpha(r)=1+\frac{1}{3}e^{-M_\psi r}$. For this metric the null tetrads in (\ref{eq8}) take the forms

\begin{equation}\label{eq19}
	\begin{aligned}
		& l^\mu=\delta_1^\mu \\
		& n^\mu=\delta_0^\mu -\frac{1}{2}\left(1-\frac{2m}{r}\alpha(r)\right) \\
		& m^\mu=\frac{1}{\sqrt{2}r}\left(\delta_2^\mu+\frac{i}{\sin\theta}\delta_3^\mu\right) \\
		& \bar{m}^\mu=\frac{1}{\sqrt{2}r}\left(\delta_2^\mu-\frac{i}{\sin\theta}\delta_3^\mu\right).
	\end{aligned}
\end{equation}

Next, the radial coordinate is replaced with complex variable as discussed in the previous section and the transformation equations (\ref{eq11}) are applied. The transformed complex null tetrads in equation (\ref{eq12}) hence, take the forms

 \begin{equation}\label{eq20}
	\begin{aligned}
		& \tilde{l}^\mu=\delta_1^\mu \\
		& \tilde{n}^\mu=\delta_0^\mu-\frac{1}{2}\left(1-\frac{2m\tilde{r}}{\rho^2}\alpha(r)\right) \\
		& \tilde{m}^\mu=\frac{1}{\sqrt{2}(\tilde{r}-ia\cos\theta)}\left[ia(\delta_0^\mu-\delta_1^\mu)\sin\theta+\delta_2^\mu+\frac{i}{\sin\theta}\delta_3^\mu\right] \\
		& \bar{\tilde{m}}^\mu=\frac{1}{\sqrt{2}(\tilde{r}+ia\cos\theta)}\left[-ia(\delta_0^\mu-\delta_1^\mu)\sin\theta+\delta_2^\mu-\frac{i}{\sin\theta}\delta_3^\mu\right].
	\end{aligned}
\end{equation}

Continuing the procedure discussed in section \ref{sec2.1}, the covariant components of the axially symmetric metric (\ref{eq17}) are derived as

 \begin{equation}\label{eq21}
g_{\mu\nu}=\left(
\begin{array}{cccc}
1-\frac{2mr}{\rho^2}\alpha(r) & 0 & 0 & \frac{2mra \sin^2\theta}{\rho^2}\alpha(r)\\
. & -\frac{\rho^2}{r^2+a^2-2mr\alpha(r)} & 0 & 0 \\
. & . & -\rho^2 & 0 \\
. & . & . & -[r^2+a^2+\frac{2mra^2\sin^2\theta}{\rho^2}\alpha(r)]\sin^2\theta
\end{array}\right).
\end{equation}

Hence, the final form of the derived metric becomes

\begin{multline}\label{eq22}
ds^2=\left[1-\frac{2mr}{\rho^2}\left(1+\frac{1}{3}e^{-M_\psi r}\right)\right]c^2dt^2
+\frac{4mra\sin^2\theta}{\rho^2}\left(1+\frac{1}{3}e^{-M_\psi r}\right)cdtd\phi\\
-\frac{\rho^2}{\Delta}dr^2-\rho^2d\theta^2
-\left[r^2+a^2+\frac{2mra^2\sin^2\theta}{\rho^2}\left(1+\frac{1}{3}e^{-M_\psi r}\right)\right]\sin^2\theta d\phi^2,
\end{multline}

where, $\rho^2=r^2+a^2\cos^2\theta$ and $\Delta=r^2-2mr\left(1+\frac{1}{3}e^{-M_\psi r}\right)+a^2$. Here, $a$ is the Kerr parameter (spin of the black hole). It is straight forward to see that the metric (\ref{eq22}) reduces to standard Kerr metric in Boyer-Lindquist coordinate system under the condition $M_\psi\rightarrow \infty$. For $a\rightarrow 0$, the spherically symmetric metric (\ref{eq2}) is retrieved. We call metric (\ref{eq22}) as Kerr-scalaron (KS) metric and will be referred to as KS metric in the rest of the paper.

\par To establish asymptotic flatness of the KS metric we proceed as follows. \citet{Kalita_2020} put forward the following relation between scalaron mass and UV and IR wave numbers ($k_{UV},k_{IR}$) of curvature corrected quantum fluctuations in the black hole background.

\begin{equation}\label{eq23}
M_\psi=\sqrt{\frac{k_{UV}^2-k_{IR}^2}{12 \ln{\left(\frac{k_{UV}}{k_{IR}}\right)}}}
\end{equation} 

Here, $k_{UV}=2\pi/\lambda_{UV}$ with $\lambda_{UV}=R_s=2m$, the event horizon and $k_{IR}=2\pi/\lambda_{IR}$ where $\lambda_{IR}$ is the IR scale associated with thermal energy density of Hawking evaporation. It is related to the Hawking temperature ($T$) as $T=\hslash c^3/8 \pi^2 K_B G M$ \citep{hawking1974black}.
Therefore, equation (\ref{eq23}) relates scalaron mass $M_\psi$ with black hole mass $m$ as 

\begin{equation}\label{eq24}
M_\psi \propto \frac{1}{m}
\end{equation}

Clearly for $r>>m$ the KS metric with leading term upto $1/r$ takes the form,

\begin{equation}\label{eq25}
ds^2=\left(1-\frac{2m}{r}\right)c^2dt^2+\frac{4ma\sin^2\theta}{r}cdt d\phi-\left(1+\frac{2m}{r}\right)dr^2-r^2\left(d\theta^2+\sin^2\theta d\phi^2\right)+ \mathcal{O}\left(\frac{1}{r^2}\right)
\end{equation} 

Equation (\ref{eq25}) confirms that the KS metric (\ref{eq22}) is asymptotically flat. This also implies that equation (\ref{eq24}) is a generic condition to ensure asymptotic flatness of the KS metric. An inverse relationship between mass of a black hole and mass of scalar field has also been obtained earlier by \citet{Carneiro:2018url} from the consideration of gravitational collapse of scalar fields leading to singularity.

\subsection{Effective Potential}\label{sec2.3}

 To study trajectory of test particles it is essential to obtain effective potential energy function. For this investigation we consider the equatorial plane $\theta=\pi/2$ for which the KS metric becomes,

\begin{equation}\label{eq26}
ds^2=\left[1-\frac{2m}{r}\alpha(r)\right]c^2dt^2+\frac{4ma}{r}\alpha(r)cdtd\phi-\frac{r^2}{\Delta}dr^2-\left[r^2+a^2+\frac{2ma^2}{r}\alpha(r)\right]d\phi^2.
\end{equation}

The covariant components of the metric (\ref{eq26}) are,

\begin{equation}\label{eq27}
\begin{aligned}
& g_{tt}=c^2\left(1-\frac{2m}{r}\alpha(r)\right) , \ \ g_{rr}=-\frac{r^2}{\Delta} , \ \  g_{t\phi}=\frac{2mac}{r}\alpha(r), \\
& g_{\phi\phi}=-\left(r^2+a^2+\frac{2ma^2}{r}\alpha(r)\right)
\end{aligned}
\end{equation}

The contravariant counterparts are worked out as (see \citet{hobson2006general} for the technique)

\begin{equation}\label{eq28}
g^{tt}=\frac{1}{c^2\Delta}\left(r^2+a^2+\frac{2ma^2}{r}\alpha(r)\right) , \ \ g^{rr}=-\frac{\Delta}{r^2} , \ \  g^{t\phi}=\frac{2ma}{cr\Delta}\alpha(r), \ \ g^{\phi\phi}=-\frac{1}{\Delta}\left(1-\frac{2m}{r}\alpha(r)\right)
\end{equation}

Since, the metric is cyclic in $t$ and $\phi$, $P_t$ and $P_\phi$ are conserved along the geodesics. As a result, equations having first integrals with respect to $t$ and $\phi$ are obtained. They are expressed as

\begin{equation}\label{eq29}
P_t=g_{tt}\dot{t}+g_{t\phi}\dot{\phi}= c^2\left(1-\frac{2m}{r}\alpha(r)\right)\dot{t}+\frac{2mac}{r}\alpha(r)\dot{\phi}=kc^2,
\end{equation}

\begin{equation}\label{eq30}
P_\phi=g_{\phi t}\dot{t}+g_{\phi \phi}\dot{\phi}=\frac{2mac}{r}\alpha(r)\dot{t}-\left(r^2+a^2+\frac{2ma^2}{r}\alpha(r)\right)\dot{\phi}=-L
\end{equation}

where, $k$ and $L$ are the constants of motion. $L$ denotes the orbital angular momentum per unit mass of the orbiting body. The expressions for $\dot{t}$ and $\dot{\phi}$ are obtained by solving the equations (\ref{eq29}) and (\ref{eq30}) and have the forms

\begin{equation}\label{eq31}
\dot{\phi}=\frac{1}{\Delta}\left[\frac{2mac}{r}\alpha(r)k+\left(1-\frac{2m}{r}\alpha(r)\right)L\right]
\end{equation}

\begin{equation}\label{eq32}
\dot{t}=\frac{1}{\Delta}\left[\left(r^2+a^2+\frac{2ma^2}{r}\alpha(r)\right)k-\frac{2ma}{rc}\alpha(r)L\right]
\end{equation} 

Using the equation $g^{\mu\nu}P_\mu P_\nu=\epsilon^2$ (where, for massive particles $\epsilon^2$ becomes $c^2$ and for a photon $\epsilon^2$ reduces to zero) gives

\begin{equation}\label{eq33}
g^{tt}(P_t)^2+2g^{t\phi}P_t P_\phi+g^{\phi\phi}(P_\phi)^2+g^{rr}(P_r)^2=\epsilon^2
\end{equation}

Here, $P_r=g_{rr}\dot{r}$. Substituting (\ref{eq28}),(\ref{eq29}) and (\ref{eq30}) in equation (\ref{eq33}) we obtain the expression for $\dot{r}$ as

\begin{equation}\label{eq34}
\dot{r}^2=c^2k^2-\epsilon^2+\frac{2\epsilon^2 m}{r}\alpha(r)+\frac{a^2(c^2k^2-\epsilon^2)-L^2}{r^2}+\frac{2m(L-ack)^2}{r^3}\alpha(r)
\end{equation}

This is the form of the energy equation in KS geometry. Now, for a massive particle ($\epsilon^2=c^2$), the energy equation is given by

\begin{equation}\label{eq35}
\frac{1}{2}\dot{r}^2+V_{\text{eff}}(r;L,k,\psi_o,M_\psi)=\frac{1}{2}c^2(k^2-1),
\end{equation}

where the effective potential is given by

\begin{equation}\label{eq36}
V_{\text{eff}}(r;L,k,\psi_o,M_\psi)=
-\frac{mc^2}{r}\alpha(r)+\frac{L^2-a^2c^2(k^2-1)}{2 r^2}-\frac{m(L-ack)^2}{r^3}\alpha(r)
\end{equation}

Now, with the redefinition $\frac{1}{2}c^2(k^2-1) = E$ where $E$ is the energy per unit mass of the orbiting body, the effective potential takes the form

\begin{multline}\label{eq37}
V_{\text{eff}}(r;L,E,\psi_o,M_\psi)=\underbrace{-\frac{mc^2}{r}+\frac{L^2}{2r^2}}_\textrm{\tiny{Newtonian Part}}\underbrace{-\frac{mL^2}{r^3}}_\textrm{\tiny{Schwarzschild Part}}
\underbrace{-\frac{a^2 E}{r^2}-\frac{ma^2(2E+c^2)}{r^3}+\frac{2maL(\sqrt{2E+c^2})}{r^3}}_\textrm{\tiny{Kerr Part}}\\
\underbrace{-\frac{mc^2}{3r}e^{-M_\psi r}}_\textrm{\tiny{Newton + Scalaron}}\underbrace{-\frac{mL^2}{3r^3}e^{-M_\psi r}}_\textrm{\tiny{Schwarzschild+Scalaron}}
\underbrace{-\frac{ma^2(2E+c^2)}{3r^3}e^{-M_\psi r}+\frac{2maL\sqrt{2E+c^2}}{3r^3}e^{-M_\psi r}}_\textrm{\tiny{Kerr + Scalaron}}
\end{multline}

It is straightforward to see that the effective potential reduces to that of standard Kerr metric when the scalaron mass becomes infinite. For $M_\psi \rightarrow \infty$ and $a \rightarrow 0$, equation (\ref{eq37}) gives the effective potential of the familiar Schwarzschild metric.

\subsection{Orbit equation}\label{2.4}

Presence of additional scalar degree of freedom near the black hole is expected to affect the orbits of test particles. We rewrite equation (\ref{eq37}) as

\begin{equation}\label{eq46}
V_{\text{eff}}(r;L,E,\psi_o,M_\psi)=-\frac{mc^2}{r}\alpha(r)+\frac{L^2}{2r^2}-\frac{mL^2}{r^3}\alpha(r)-\frac{a^2E}{r^2}
-\frac{ma^2(2E+c^2)}{r^3}\alpha(r)+\frac{2maL\sqrt{2E+c^2}}{r^3}\alpha(r)
\end{equation}

Also, equation (\ref{eq35}) can be expressed as,

\begin{equation}\label{eq47}
\frac{1}{2}\dot{r}^2+V_{\text{eff}}(r;L,E,\psi_o,M_\psi)=E
\end{equation}

Now, $\dot{r}$ can be redefined as $\dot{r}=\dot{\phi}r'$, where, $r'=\frac{\partial r}{\partial \phi}$. In terms of $r'$, equation (\ref{eq47}) can be expressed as

\begin{equation}\label{eq48}
r'^2=2 \ \frac{[E-V_{\text{eff}}(r;L,E,\psi_o,M_\psi)]}{\dot{\phi}^2}
\end{equation}

With equation (\ref{eq31}) it becomes

\begin{equation}\label{eq49}
r'^2=2 \ \frac{[E-V_{\text{eff}}(r;L,E,\psi_o,M_\psi)]}{\frac{1}{\Delta^2}\left[\frac{2mac}{r}\alpha(r)k+\left(1-\frac{2m}{r}\alpha(r)\right)L\right]^2}
\end{equation}

Substituting equation (\ref{eq46}) in (\ref{eq49}) and expanding in the first order of $a$, the expression reduces to

\begin{equation}\label{eq50}
r'^2\approx \left(\frac{2Er^4+2m\alpha(r)c^2r^3-r^2L^2+2m\alpha(r)L^2r}{L^2}\right)+\left(\frac{8m\alpha(r)r^3\sqrt{2E+c^2}(m\alpha(r)c^2+Er)}{L^3(2m\alpha(r)-r)}\right)a+\mathcal{O}(a^2)
\end{equation} 

Now, the term $\frac{1}{L^3(2m\alpha(r)-r)}$ in (\ref{eq50}) is expressed as $\frac{1}{L^3(2m\alpha(r)-r)}\approx-\frac{1}{L^3 r}\left(1+\frac{2m\alpha(r)}{r}\right)$. Substituting this expansion in equation (\ref{eq50}) and considering the first order terms in $r$,
the equation reduces to

\begin{equation}\label{eq51}
r'^2\approx\frac{2Er^4}{L^2}+\frac{2mc^2r^3}{L^2}\alpha(r)-r^2+2mr\alpha(r)-\frac{8a\sqrt{2E+c^2}m^2c^2r^2}{L^3}\alpha(r)^2-\frac{8aE\sqrt{2E+c^2}mr^3}{L^3}\alpha(r)
\end{equation}

Defining $u=1/r \implies u'=-\frac{1}{r^2}r'$ we obtain the differential equation of the orbit as

\begin{multline}\label{eq52}
u''+u=\frac{mc^2}{L^2}\alpha(u)+3m\alpha(u)u^2-\frac{8a\sqrt{2E+c^2}m^2c^2}{L^3}\alpha(u)^2u-\frac{4aE\sqrt{2E+c^2}m}{L^3}\alpha(u)+\frac{mc^2M_\psi}{3L^2}u^{-1}e^{-\frac{M_\psi}{u}}+\frac{mM_\psi}{3}ue^{-\frac{M_\psi}{u}}\\
-\frac{8a\sqrt{2E+c^2}m^2c^2M_\psi}{3L^3}\alpha(u)e^{-\frac{M_\psi}{u}}-\frac{4aE\sqrt{2E+c^2}mM_\psi}{3L^3}u^{-1}e^{-\frac{M_\psi}{u}}
\end{multline}

Here, $\alpha(u)=1+\frac{1}{3}e^{-\frac{M_\psi}{u}}$. Equation (\ref{eq52}) represents the equation of motion of a test particle orbiting the KS black hole. Due to the cumbersome nature of the differential equation, it is not possible to apply standard perturbative methods to solve the equation. But, when the scalaron mass becomes infinite, the equation reduces to the standard equation of motion for Kerr black hole. Further, when $M_\psi\rightarrow \infty,\ a\rightarrow 0$, it becomes the equation of motion for a test particle near a Schwarzschild black hole. 

\subsection{Pericenter shift}\label{2.5}

The advance in periapse of a test particle near the black hole should naturally undergo some modification due to the presence of scalarons. Presence of exponential terms along with the algebraic terms in (\ref{eq52}), makes it challenging to identify the secular term in the non-linear differential equation that is responsible for the precession of the orbits. Hence, we resort to an alternative to investigate this effect. Following the technique given in \citet{rovelli2021general} an expression for the angular frequency and radial frequency of the orbit of the test particle is derived. The test particle is considered to be slowly orbiting the central mass with $E<<c^2$. The local minimum of the potential (\ref{eq37}) at radius $r=r_{*}$ (where, $r_*=a'(1-e^2)$ with $a'$ being the semi-major axis) is obtained from

\begin{equation}\label{eq38}
\frac{d V_{\text{eff}}}{dr}\bigg|_{r=r_*}=0
\end{equation}

This leads to the condition,

\begin{multline}\label{eq39}
\frac{mc^2}{r_*^2}\left(1+\frac{1}{3}e^{-M_\psi r_*}\right)=\frac{L^2}{r_*^3}-\frac{3mL^2}{r_*^4}\left(1+\frac{1}{3}e^{-M_\psi r_*}\right)-\frac{2 a^2 E}{r_*^3}-\frac{3 ma^2(2E+c^2)}{r_*^4}\left(1+\frac{1}{3}e^{-M_\psi r_*}\right)\\
+\frac{6maL\sqrt{2E+c^2}}{r_*^4}\left(1+\frac{1}{3}e^{-M_\psi r_*}\right)
-\frac{mc^2}{3 r_*}M_\psi e^{-M_\psi r_*}-\frac{mL^2}{3 r_*^3}M_\psi e^{-M_\psi r_*}\\
-\frac{ma^2(2E+c^2)}{3r_*^3}M_\psi e^{-M_\psi r_*}+\frac{2maL\sqrt{2E+c^2}}{3r_*^3}M_\psi e^{-M_\psi r_*}
\end{multline}

The second derivative of the effective potential is equivalent to square of the radial frequency ($\omega_r^2$) of the orbit of the test particle. After some rigorous algebra, an expression for $\omega_r^2$ is obtained as,

\begin{equation}\label{eq40}
\omega_r^2=\frac{d^2V_{\text{eff}}}{dr^2}\bigg|_{r=r_*}=G(r_*)
\end{equation}

where, after applying the condition $E<<c^2$ , 

\begin{multline}\label{eq41}
G(r_*)=\frac{L^2}{r_*^4}-\frac{6mL^2}{r_*^5}\left(1+\frac{1}{3}e^{-M_\psi r_*}\right)-\frac{6ma^2c^2}{r_*^5}\left(1+\frac{1}{3}e^{-M_\psi r_*}\right)+\frac{12maLc}{r_*^5}\left(1+\frac{1}{3}e^{-M_\psi r_*}\right)-\frac{4mL^2}{3r_*^4}M_\psi e^{-M_\psi r_*}\\
-\frac{4ma^2c^2}{3r_*^4}M_\psi e^{-M_\psi r_*}+\frac{8maLc}{3r_*^4}M_\psi e^{-M_\psi r_*}-\frac{mc^2}{3r_*}M_\psi^2 e^{-M_\psi r_*}\\
-\frac{mL^2}{3 r_*^3}M_\psi^2 e^{-M_\psi r_*}-\frac{ma^2c^2}{3 r_*^3} M_\psi^2 e^{-M_\psi r_*}+\frac{2maLc}{3r_*^3}M_\psi^2 e^{-M_\psi r_*}
\end{multline}

The angular frequency of the orbit of the test particle is given by $\omega_\phi=\dot{\phi}$ . Thus, using equation (\ref{eq31}) and applying $E<<c^2$,

\begin{equation}\label{eq42}
\omega_\phi^2= H(r_*)
\end{equation}

where,

\begin{equation}\label{eq43}
H(r_*)=\frac{1}{\Delta^2 r_*^2}\left[2mac\left(1+\frac{1}{3}e^{-M_\psi r_*}\right)+\left(r_*-2m\left(1+\frac{1}{3}e^{-M_\psi r_*}\right)\right)L\right]^2
\end{equation}

From the equations (\ref{eq40}) and (\ref{eq42}), it is clear that the degeneracy between angular frequency and radial frequency is broken. Hence, the orbit undergoes precession by an amount, $\delta\phi=\phi-2\pi$. This shift occurs in the time period, $T_r=2\pi/\omega_r$. The radial frequency ($\omega_r$) can be expressed in terms of angular frequency ($\omega_\phi$) as, (after expanding in the first order)

\begin{equation}\label{eq44}
\omega_r=\omega_\phi\left[1-\frac{1}{2}\left(1-\frac{G(r_*)}{H(r_*)}\right)\right]
\end{equation}

Using the relation $\phi=\omega_\phi T_r$ and again expanding upto first order, we get the precession as,

\begin{equation}\label{eq45}
\delta\phi=2\pi\left[\frac{1}{2}\left(1-\frac{G(r_*)}{H(r_*)}\right)\right]
\end{equation}

Equation (\ref{eq45}) represents an approximate form of the pericenter shift in KS geometry. It reduces to GR pericentre shift  ($\delta\phi_{GR}=\frac{6\pi m}{r_*}$) as soon as $M_\psi\rightarrow \infty$ and $a\rightarrow0$ (Schwarzschild).

\subsection{Black hole shadow in KS-metric}\label{sec2.6}

The horizon scale images of the GC black hole, Sgr A*, extracted by \citet{Event_Horizon_Telescope_Collaboration_2022} , offer a probe to inspect the strong field regime of gravity.  The current image of the GC black hole taken by the EHT consists of a bright emission ring, surrounding a central brightness depression (the black hole shadow). The angular diameter of the shadow has been reported as $d_{sh} = 48.7 \pm 7.0$ $\mu$as \citep{Akiyama_VI}. This scale has been taken into account in order to constrain the metric under consideration.
\par The Schwarzschild black hole when observed from infinity has a photon capture radius of $3\sqrt 3m$ \citep{Akiyama_VI}. The scale of the event horizon is $R_s = 2m$. Photons having impact parameter $b < 3\sqrt 3 m$ plunge into the black hole, while those with $b > 3\sqrt3 m$ escape to infinity. The photons with $b = 3\sqrt 3 m$ are trapped into an unstable orbit. Thus the occurrence of a central brightness depression of size $3\sqrt3 m$ constitutes the black hole shadow. For Kerr black holes, the radii of photon orbits have a strong dependence on spin, but a coincidental cancellation of the frame dragging effects and quadrupole structure of the Kerr metric causes the black hole shadow to be very weakly dependent on spin \citep{Johannsen_2010}. The shadow scale is calculated as follows. A general static and spherically symmetric metric has the form, \citep{kalita2023constraining}

\begin{equation}\label{eq56}
ds^2 = g_{tt}(r, P_i)c^2dt^2 + \Sigma_{i,j = 1,2,3}g_{ij}(r, P_i)dx^idx^j
\end{equation}

where $P_i$ are the parameters of the metric. The above choice is taken into consideration as the size of the black hole shadow is very weakly depended on the black hole spin. \citep{Johannsen_2010,Johannsen_2013}.\\
The shadow radius is given by\citep{Akiyama_VI}

\begin{equation}\label{eq57}
r_{sh} = \frac{r_{ph}(P_i)}{\sqrt{g_{tt}(r_{ph}, P_i)}}
\end{equation}

In the above equation $r_{ph}$ denotes the photon radius which is given by the solution of the implicit equation

\begin{equation}\label{eq58}
r_{ph}(P_i) = 2g_{tt}(r_{ph}, P_i)\left(\frac{dg_{tt}(r, P_i)}{dr}\right)_{r_{ph}}^{-1}
\end{equation}

By solving the above implicit equation, and substituting the value in (\ref{eq57}), the parameters of the metric, $P_i$ can be extracted from the measured angular diameter of the black hole shadow. A prior value of distance to the black hole (D) is required for producing constraints on the parameters $P_i$ from the relation for angular size of the shadow, $d_{sh}$ = $2r_{ph}(P_i)/D$.

\section{Astronomical Prospects}\label{sec3}

\subsection{Scalaron mass from the shadow of Sgr A*}\label{sec3.1}

In this section we investigate scalaron mass for which the KS metric becomes compatible with the observed angular diameter of the shadow of Sgr A*. The $g_{tt}$ component of the KS metric (\ref{eq22}) has the form

\begin{equation}\label{eq59a}
g_{tt} = 1 - \frac{2mr}{r^2+a^2cos^2\theta}-\frac{2mre^{-M_{\psi}r}}{3(r^2+a^2cos^2\theta)}
\end{equation}

The above expression is substituted in equation (\ref{eq58}) to obtain the implicit equation for the photon radius in KS metric. It has the form

\begin{multline}\label{eq59}
3r_{ph}^4 - M_{\psi}mr_{ph}^4e^{-M_{\psi}r_{ph}}-9mr_{ph}^3-3mr_{ph}^3e^{-M_{\psi}r_{ph}}
-M_{\psi}ma^2cos^2\theta r_{ph}^2e^{-M_{\psi}r_{ph}}+6r_{ph}^2a^2cos^2\theta\\
-3mr_{ph}a^2cos^2\theta+ma^2cos^2\theta r_{ph}e^{-M_{\psi}r_{ph}}-2mr_{ph}e^{-M_{\psi}r_{ph}}a^2cos^2\theta+3a^4cos^4\theta = 0
\end{multline}

The spin parameter $a$ is parameterized as $a = m\chi$ where $ \chi\leq\ 1$. In a recent study, the spin of the GC black hole has been reported as $\chi=0.90\pm 0.06$ \citep{10.1093/mnras/stad3228}. Hence, $\chi= 0.90$ has been chosen. In terms of the spin parameter $\chi$, equation (\ref{eq59}) becomes

\begin{multline}\label{eq60}
3r_{ph}^4 - M_{\psi}mr_{ph}^4e^{-M_{\psi}r_{ph}}-9mr_{ph}^3-3mr_{ph}^3e^{-M_{\psi}r_{ph}}-M_{\psi}m(m\chi)^2cos^2\theta r_{ph}^2e^{-M_{\psi}r_{ph}}+6r_{ph}^2(m\chi)^2cos^2\theta \\
-3mr_{ph}(m\chi)^2cos^2\theta+m(m\chi)^2cos^2\theta r_{ph}e^{-M_{\psi}r_{ph}}-2mr_{ph}e^{-M_{\psi}r_{ph}}(m\chi)^2cos^2\theta+3(m\chi)^4cos^4\theta = 0
\end{multline}

The above equation reduces to the implicit equations for the Kerr metric if $M_{\psi} \rightarrow \infty$ and Schwarzschild metric if $M_{\psi} \rightarrow \infty$ and $a\rightarrow 0$ as obtained by \citet{kalita2023constraining}, which have the following forms

\begin{equation}\label{eq61}
\begin{aligned}
& r_{ph}^{4}-3mr_{ph}^{3}+2r_{ph}^{2}(\chi m)^{2}cos^{2}\theta - mr_{ph}(\chi m)^{2}cos^{2}\theta+(\chi m)^{4}cos^{4}\theta =0 \ \ \ \text{(Kerr)}\\
& r_{ph}^{4}-3mr_{ph}^{3} =0\ \ \ \text{(Schwarzschild)}
\end{aligned}
\end{equation}

Clearly $r_{ph} = 3m$ for the Schwarzschild case. As $g_{tt} = 1-\frac{2m}{r}$ for the Schwarzschild metric equation (\ref{eq57}) produces the shadow radius as $r_{sh} = 3\sqrt3 m$, which is a known result.

\par Equation (\ref{eq60}) has been solved for $\chi=0.90$  and inclination angles $\theta=0^o$ \& $90^o$. For Sgr A*, $m=GM/c^2=0.042$ au. The scalaron mass is considered in the range $10^{-5}$ au$^{-1}$ ($10^{-22}$ eV) - $12.22$ au$^{-1}$ ($10^{-16}$ eV)(see \citet{Kalita_2020} for comparison). The shadow diameter for these range of scalaron mass is obtained for the two inclination angles mentioned above. The distance to the GC black hole is taken as $D\approx8$ Kpc \citep{2019A&A...625L..10G}. It is to be noted that the angular size of shadow is insensitive to the spin of the black hole and hence is not a very strong probe for constraining spacetime metric. However, the angular size of the emission ring can be used to constrain gravity theory or the metric \citep{Akiyama_VI}. Diameter of the emission ring ($d_{\text{e-ring}}$), can be used to measure the properties of the black hole metric and to access its compatibility with Kerr solution in GR for a black hole of given angular size. The relation between the diameter of emission ring and the corresponding shadow size ($d_{sh}$) is expressed as,\citep{Akiyama_VI}

\begin{equation}\label{eq68}
    d_{\text{e-ring}}=\frac{d_{\text{e-ring}}}{d_{sh}}d_{sh}=\alpha_c(1+\delta)d_{sh,Sch}
\end{equation}

where, $\alpha_c=\frac{d{\text{e-ring}}}{d_{sh}}$ is the $\alpha$-calibration factor. It measures the correlation between the scales of emission ring and shadow. For GC black hole, $d_{\text{e-ring}}$ has been reported to be $51.8\pm 2.3\ \mu$as \citep{Event_Horizon_Telescope_Collaboration_2022}. The parameter $\delta$ is expressed as,

\begin{equation}\label{eq67}
    \delta=\frac{d_{sh}}{d_{sh,Sch}}-1
\end{equation}

It measures the deviation of the estimated shadow diameter from that of a Schwarzschild black hole ($d_{sh,Sch}=6\sqrt{3}\theta_g$, where $\theta_g$ is the angular size of half of the gravitational radius). \citet{Event_Horizon_Telescope_Collaboration_2022} constrained $\delta$ as $-0.08_{-0.09}^{+0.09}$ and $-0.04_{-0.10}^{+0.09}$ for GC black hole using VLTI and Keck data for mass and distance to the black hole respectively. It is also to be noted that the black hole is compatible with Kerr metric if $\delta$ is in the range $[-0.075,0]$ \citep{Akiyama_VI}. For GC black hole, $d_{sh,Sch}$ is estimated to be $55.1107\ \mu$as. Using equation (\ref{eq67}), $\delta$ is estimated for two models: SchS metric and KS metric. Also, for estimation of calibration factor $\alpha_c$, $d_{\text{e-ring}}$ is chosen as $51.8\ \mu$as. The estimated values of $\delta$ for SchS metric is presented in Table \ref{table1}. The estimated values of $\delta$ for KS metric with spin $\chi=0.90$ are  shown in Table \ref{table3}.

\begin{table}[hptb]
\centering
\caption{Estimated values of deviation parameter $\delta$ in SchS geometry for different scalaron masses}
\label{table1}
\begin{tabular}{cccc}
\hline
\begin{tabular}[c]{@{}c@{}}Scalaron Mass,\\ $M_\psi$ (eV)\end{tabular} & \begin{tabular}[c]{@{}c@{}}$d_{sh}$\\ ($\mu$as)\end{tabular} & $\alpha_c$ & $\delta$ \\ \hline
$1.01 \times 10^{-22}$ & $73.4809$ & $0.704945$ & $0.33333$ \\
$1.01 \times 10^{-21}$ & $73.4805$ & $0.704948$ & $0.33332$ \\
$1.01 \times 10^{-20}$ & $73.4771$ & $0.704980$ & $0.33326$ \\
$1.01 \times 10^{-19}$ & $73.4432$ & $0.705306$ & $0.33265$ \\
$1.01 \times 10^{-18}$ & $73.1091$ & $0.708530$ & $0.32658$ \\
$1.01 \times 10^{-17}$ & $70.1879$ & $0.738018$ & $0.27358$ \\
$5.13 \times 10^{-17}$ & $62.5412$ & $0.828252$ & $0.13483$ \\
$9.25 \times 10^{-17}$ & $59.0488$ & $0.877239$ & $0.07145$ \\
$1.01 \times 10^{-16}$ & $58.5922$ & $0.884076$ & $0.06317$ \\ \hline
\end{tabular}
\end{table}

\begin{table}[hptb]
\caption{Estimated values of deviation parameter $\delta$ in KS geometry for different scalaron masses for $\chi=0.90$}
\label{table3}
\begin{tabular}{ccccccc}
\hline
\multirow{2}{*}{\begin{tabular}[c]{@{}c@{}}Scalaron Mass,\\ $M_\psi$ (eV)\end{tabular}} &
  \multicolumn{3}{c}{$\theta=0^o$} &
  \multicolumn{3}{c}{$\theta=90^o$} \\ \cline{2-7} 
 &
  \begin{tabular}[c]{@{}c@{}}$d_{sh}$\\ ($\mu$as)\end{tabular} &
  $\alpha_c$ &
  $\delta$ &
  \begin{tabular}[c]{@{}c@{}}$d_{sh}$\\ ($\mu$as)\end{tabular} &
  $\alpha_c$ &
  $\delta$ \\ \hline
$1.01 \times 10^{-22}$ & $69.3801$ & $0.746610$ & $0.25892$  & $73.5153$  & $0.704614$ & $0.333957$ \\
$1.01 \times 10^{-21}$ & $69.3798$ & $0.746614$ & $0.25891$  & $73.5149$ & $0.704618$ & $0.333951$ \\
$1.01 \times 10^{-20}$ & $69.3765$ & $0.746649$ & $0.25885$  & $73.5114$ & $0.704652$ & $0.333887$ \\
$1.01 \times 10^{-19}$ & $69.3437$ & $0.747002$ & $0.25826$  & $73.4787$ & $0.704965$ & $0.333293$ \\
$1.01 \times 10^{-18}$ & $69.0201$ & $0.750505$ & $0.25239$  & $73.1442$ & $0.708189$  & $0.327224$ \\
$1.01 \times 10^{-17}$ & $66.1678$ & $0.782857$ & $0.20063$  & $70.2233$ & $0.737646$  & $0.274222$ \\
$5.13 \times 10^{-17}$ & $58.4112$ & $0.886815$ & $0.05988$  & $62.5824$ & $0.827708$ & $0.135577$ \\
$9.25 \times 10^{-17}$ & $54.5872$ & $0.948939$ & $\mathbf{-0.00949}$ & $59.0968$ & $0.876527$ & $0.072330$ \\
$1.01 \times 10^{-16}$ & $54.0584$ & $0.958222$ & $\mathbf{-0.01909}$ & $58.6427$ & $0.883314$  & $0.064090$ \\ \hline
\end{tabular}
\end{table}

It is observed that for SchS metric, the deviation parameter for all scalaron masses is outside the bound reported by EHT observations. On the other hand, for KS metric with $\chi=0.90$ and $\theta=0^o$ the deviation parameter for massive scalarons ($10^{-17}$ eV \& $10^{-16}$ eV),is within the EHT bounds. The values of deviation parameter which are consistent with Kerr metric are marked as bold in Table \ref{table3}.

\subsection{Schwarzschild pericenter shift}\label{sec3.2}

Pericenter shift of keplerian orbit is a strong probe for constraining gravitation theory.The Schwarzschild precession of $12'$ per orbital period in S2's orbit has been measured by \citet{gravity2020}. Here we take the Schwarzschild limit of pericenter shift in KS metric by setting $a=0$ in equation \ref{eq45}. The resulting expression is written as,\footnote{using $L^2=mc^2 a'(1-e^2)$ and $r_*=a'(1-e^2)$.}

\begin{multline}\label{eq62}
    (\delta\phi)_{SchS}=\frac{6\pi m}{a'(1-e^2)}\left(1+\frac{1}{3}e^{-M_\psi a'(1-e^2)}\right)+\frac{4\pi m M_\psi}{3}e^{-M_\psi a'(1-e^2)}\\
    +\frac{2\pi a'^2(1-e^2) M_\psi^2}{6}e^{-M_\psi a'(1-e^2)}+\frac{2\pi m a'(1-e^2) M_\psi^2}{6}e^{-M_\psi a'(1-e^2)}
\end{multline}

Here, $(\delta\phi)_{SchS}$ represents precession in SchS geometry. Deviation from standard (GR based) Schwarzschild precession($(\delta\phi)_{Sch}=\frac{6\pi m}{a'(1-e^2)}$) is obtained as

\begin{multline}\label{eq63}
    (\delta\phi)_{SchS}-(\delta\phi)_{Sch}=\frac{6\pi m}{3 a'(1-e^2)}e^{-M_\psi a'(1-e^2)}+\frac{4\pi m M_\psi}{3}e^{-M_\psi a'(1-e^2)}\\
     +\frac{2\pi a'^2(1-e^2) M_\psi^2}{6}e^{-M_\psi a'(1-e^2)}+\frac{2\pi m a'(1-e^2) M_\psi^2}{6}e^{-M_\psi a'(1-e^2)}
\end{multline}

\begin{figure}
    \centering
    \includegraphics[width=\columnwidth]{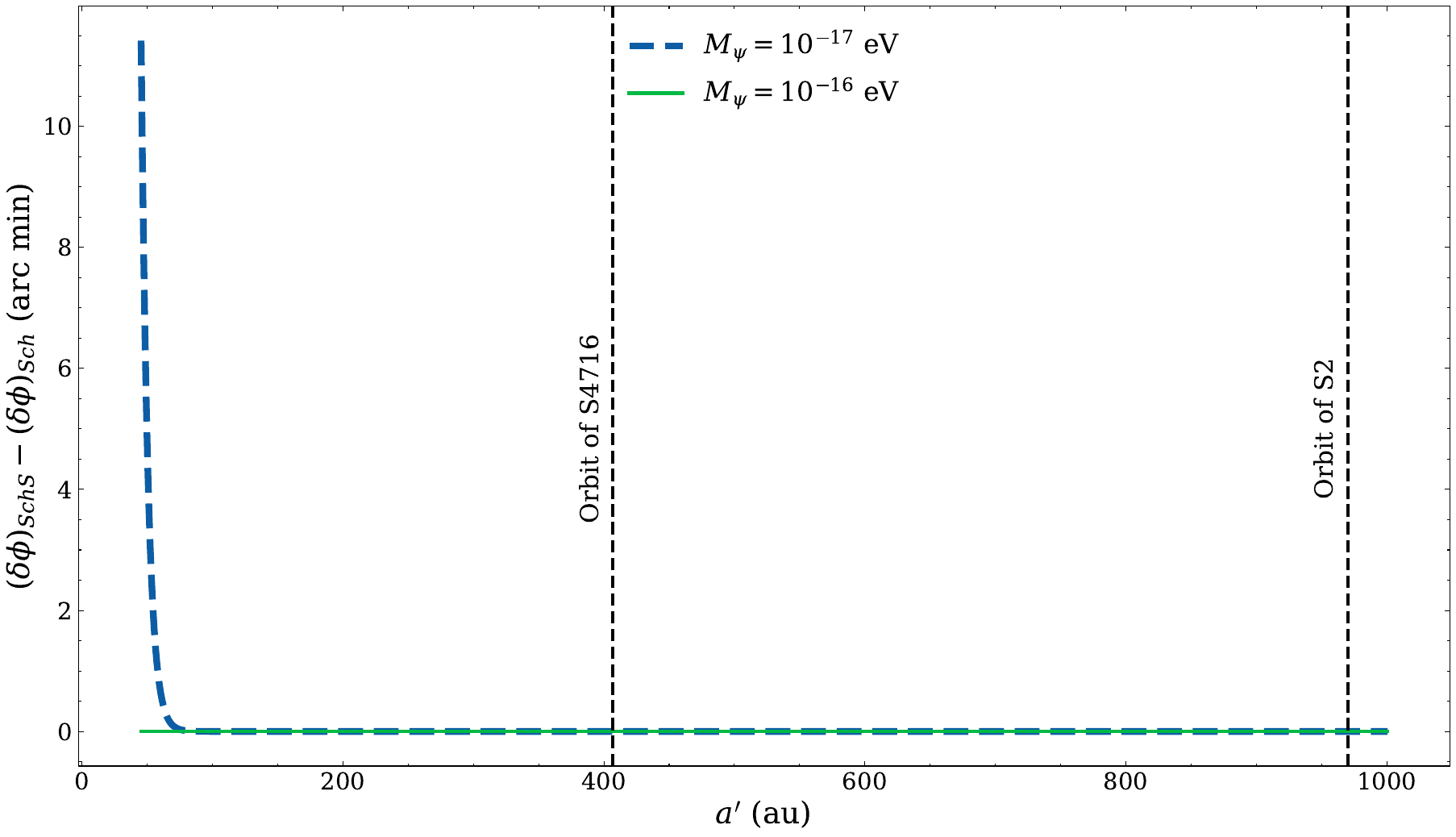}
    \caption{Deviation of Schwarzschild-scalaron pericenter from standard Schwarzschild pericenter shift with respect to semi-major axis ($a'$) for two scalaron masses. The orbital size of the stars S4716 and S2 are shown by two vertical dashed lines in the figure.}
    \label{fig4}
\end{figure}

The deviation is studied against semi-major ($a'$) for scalarons with $M_\psi=10^{-17}$ eV - $10^{-16}$ eV constrained by the shadow of Sgr A*. We take a minimum bound on $a'$ which is estimated from the consideration that time scale of gravitational wave emission from a star orbiting the GC black hole must be greater than or equal to the minimum age of the star ($\sim 600$ Myr for most of the S-stars) encircling the GC black hole (see \citet{doi:10.1142/S0218271823500219} and references therein). For S2 like eccentricity ($e=0.9$), this value is obtained as $45.40$ au. We study the deviation upto S2 like orbit with $a'=1000$ au. The variation is presented in Figure \ref{fig4}. 

\par It is observed that $10^{-16}$ eV scalarons almost produce GR like pericenter shift (see the flat graph in figure \ref{fig4}) at all orbits. But $10^{-17}$ eV scalarons dominate general relativistic precession for very compact orbits ($a'<<50$ au). For orbits of known S-stars such as S2 and the recently discovered S4716 \citep{Peißker_2022} both these scalarons show GR like precession. This result is in harmony with the previous result (section \ref{sec3.1}) that that massive scalarons reproduce GR like shadow.

\par The Schwarzschild precession of orbits near the GC black hole is parameterized as \citep{gravity2020}

\begin{equation}\label{eq64}
    \delta\phi=f_{SP} \times (\delta\phi)_{Sch}
\end{equation}

where, $f_{SP}$ measures departure from the standard Schwarzschild precession. $f_{SP}=0$ and $1$ for Newtonian gravity and GR respectively. For scalarons we have

\begin{equation}\label{eq65}
    f_{SP}=1+\left[\frac{(\delta\phi)_{SchS}-(\delta\phi)_{Sch}}{(\delta\phi)_{Sch}}\right]
\end{equation}

The \citet{gravity2020} has constrained the $f_{SP}$ parameter from the precession of the orbit of S2 as $f_{SP}=1.10\pm 0.29$. Here, the deviation $(\delta\phi)_{SchS}-(\delta\phi)_{Sch}$ has been estimated from Figure \ref{fig4} for three hypothetical orbits near the GC black hole having semi-major axes $45$ au, $50$ au and $100$ au. The Schwarzschild precession for the three orbits has been estimated for S2-like eccentricity ($e=0.9$). For $45$ au orbit, the deviation is estimated as 11.57 arc minute. The scalaron mass required for the calculation is chosen as $10^{-17}$ eV. The estimated $f_{SP}$ values are shown in Table \ref{tab1}. It is seen that the KS metric with $10^{-17}$ eV scalarons is compatible with GR.

\begin{table}[hptb]
\centering
\caption{$f_{SP}$ values for three hypothetical orbits along with the constraint on $f_{SP}$ obtained from S2's orbit}
\label{tab1}
\begin{tabular}{ccc}
\hline
Stellar orbits & Semi-major axis ($a'$) & $f_{SP}$ \\ \hline
S2's orbit & $970$ au & \begin{tabular}[c]{@{}c@{}}$1.10 \pm 0.29$\\ \citep{gravity2020}\end{tabular} \\ \hline
\multirow{3}{*}{Hypothetical stellar orbits}                 & $100$ au & $1.00$    \\
               & $50$ au                & $1.02$  \\
               & $45$ au                & $1.04$  \\ \hline
\end{tabular}
\end{table}

\subsection{Lens-Thirring effect in KS geometry}\label{sec3.2b}

Any test gyroscope in a stationary spacetime is known to exhibit the Lens–Thirring (LT) precession effect, which causes dragging of locally inertial frames along the rotating spacetime \citep{lense1918influence}. In the weak field limit the LT precession frequency is expressed as \citep{hartle}

\begin{equation}\label{eq69}
    \vec{\Omega}_{LT}=\frac{G}{c^2r^3}\left[3(\Vec{J}.\hat{r})\hat{r}-\vec{J}\right]
\end{equation}

where, $\hat{r}$ is the unit vector along r and $J=Mac$ is the angular momentum. This is the particular case of a more general phenomenon known as frame dragging. Plane of a stellar orbit(a gyroscope) surrounding a rotating black hole undergoes precession due to this effect \citep{Will_2008}. With the KS metric we obtain the LT frequency without using any weak field approximation. We adopt the method given by \citet{Chakraborty_2014}. The LT frequency (in vector form) for KS metric obtained in the present study is expressed as

\begin{equation}\label{eq70}
    \vec{\Omega}^{KS}_{LT}=2mac\alpha(r) \cos\theta\frac{r\sqrt{\Delta}}{\rho^3[\rho^2-2mr\alpha(r)]}\hat{r}-mac \sin\theta\frac{[\rho^2(\alpha(r)-\frac{M_\psi r}{3}e^{-M_\psi r})-2r^2 \alpha(r)]}{\rho^3[\rho^2-2mr\alpha(r)]}\hat{\theta}
\end{equation}

The magnitude of (\ref{eq70}) is expressed as

\begin{equation}\label{eq71}
    \Omega^{KS}_{LT}=\frac{mac}{\rho^3[\rho^2-2mr\alpha(r)]}\left[4\alpha(r)^2 r^2 \Delta \cos^2\theta+\left(\rho^2\left(\alpha(r)-\frac{M_\psi r}{3}e^{-M_\psi r}\right)-2r^2\alpha(r)\right)^2\sin^2\theta\right]^{1/2}
\end{equation}

\begin{figure*}
\begin{center}
    \begin{minipage}[b]{0.48\columnwidth}
        \includegraphics[width=\columnwidth]{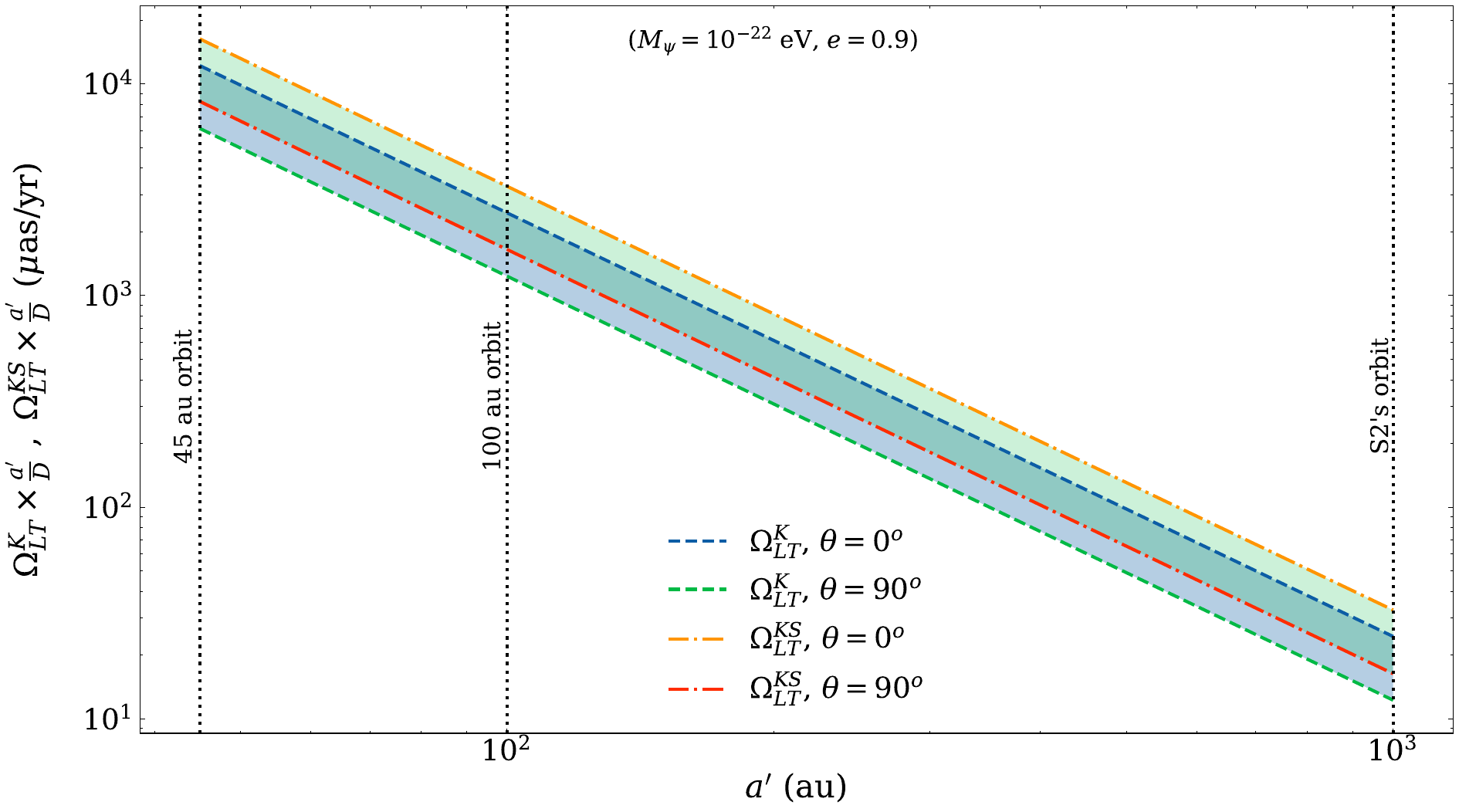}
    \end{minipage}
    \begin{minipage}[b]{0.48\columnwidth}
        \includegraphics[width=\columnwidth]{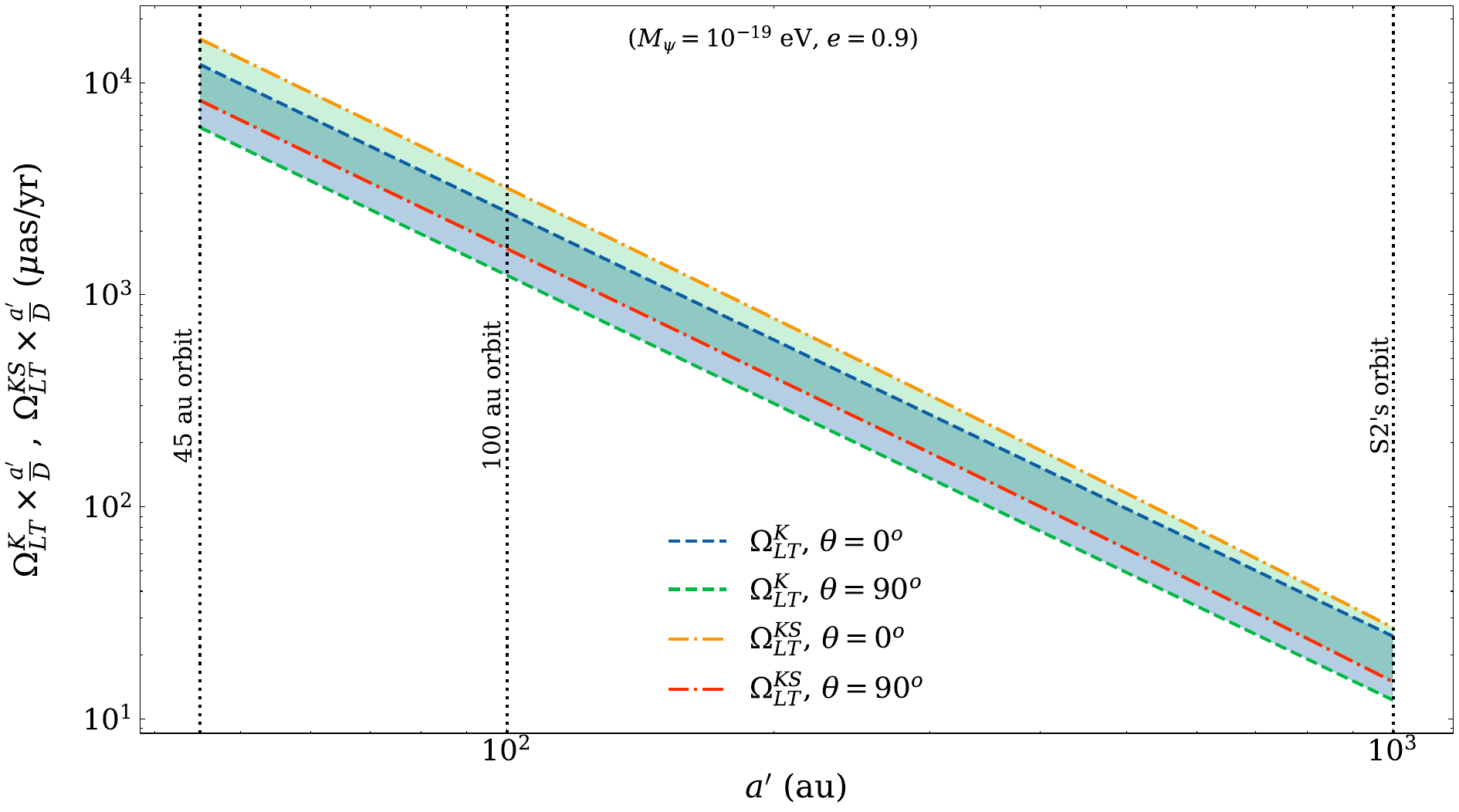}
    \end{minipage}\\ \vspace{0.3cm}
    \begin{minipage}[b]{0.48\columnwidth}
        \includegraphics[width=\columnwidth]{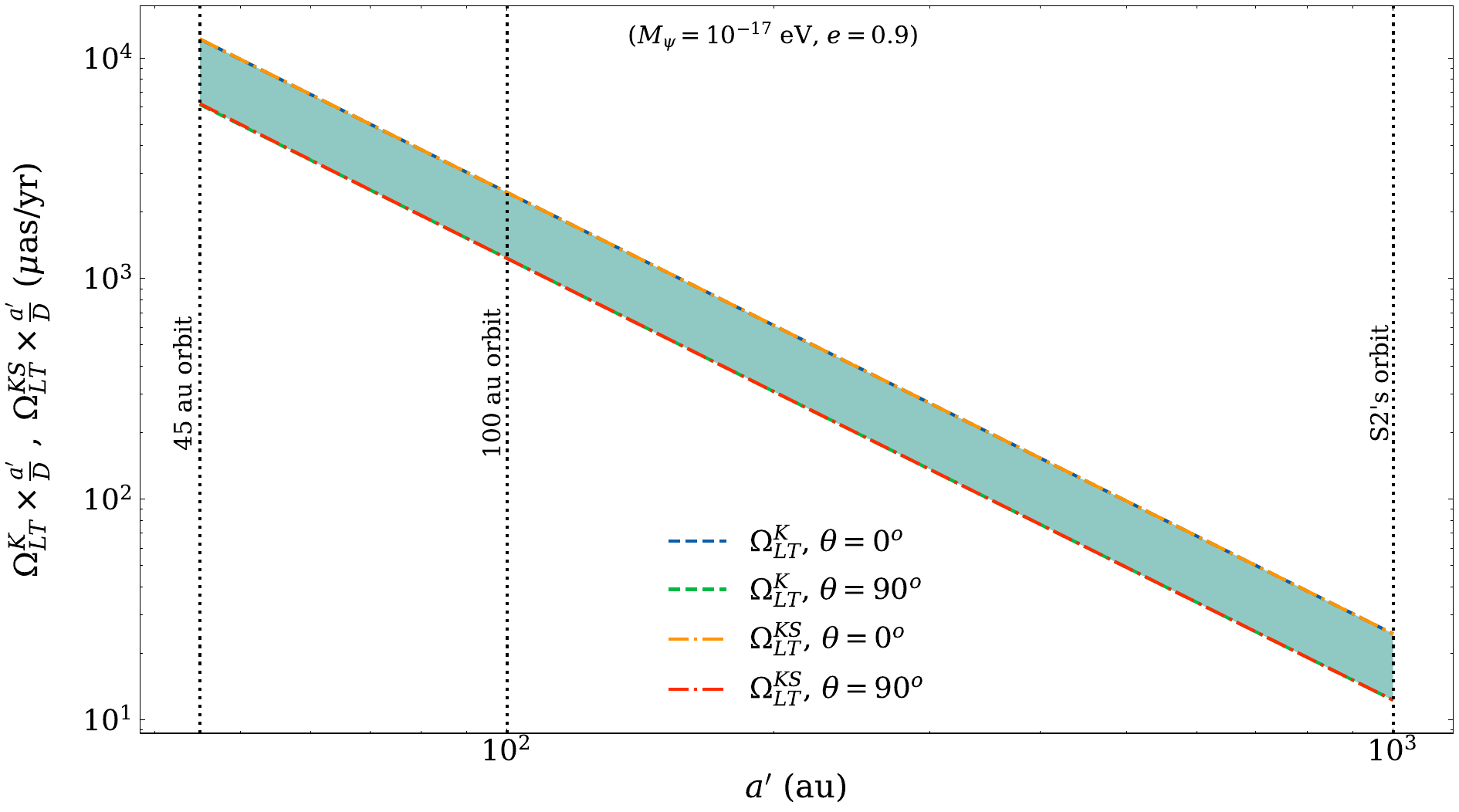}
    \end{minipage}
    \begin{minipage}[b]{0.48\columnwidth}
        \includegraphics[width=\columnwidth]{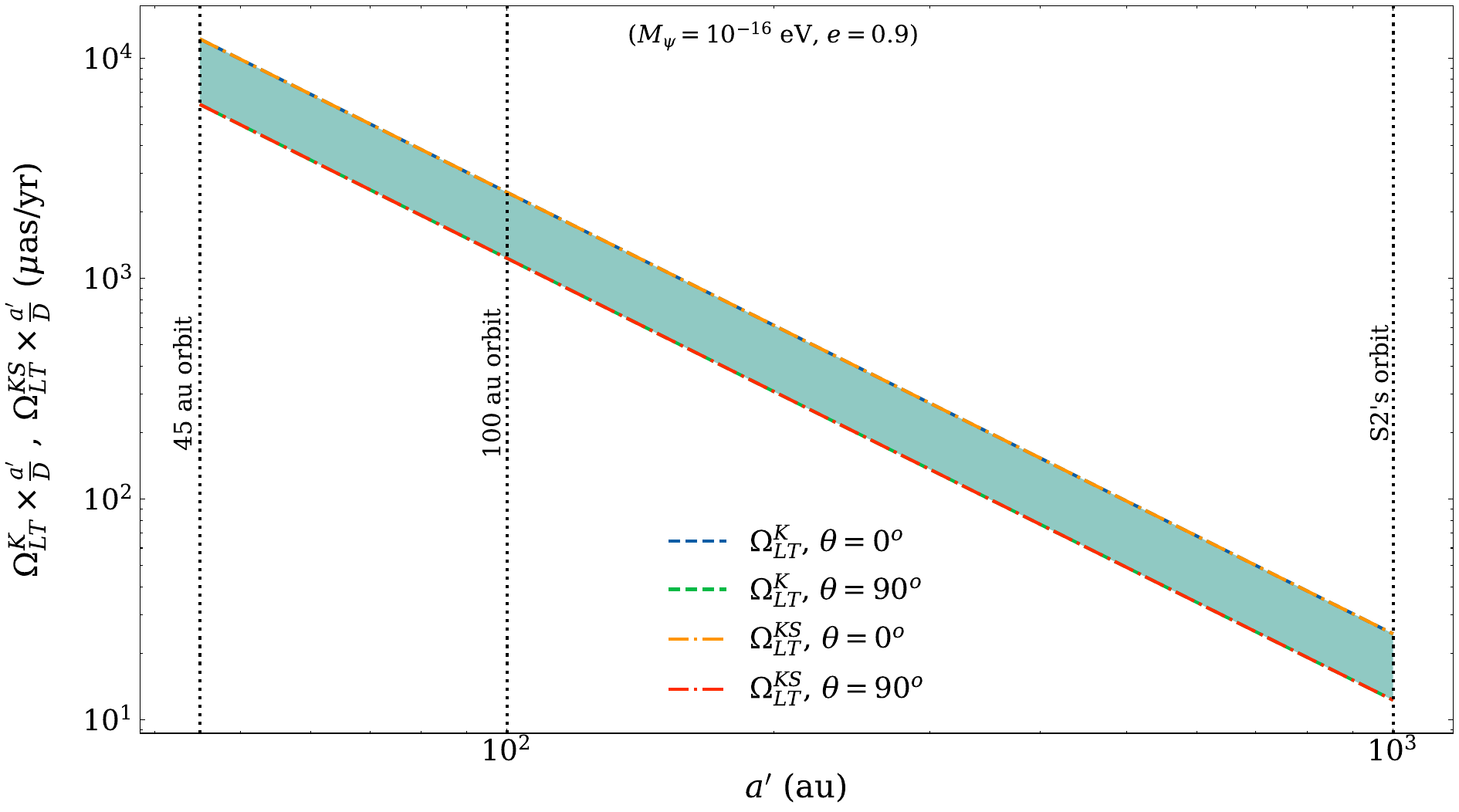}
    \end{minipage}
\end{center}
\caption{The variation of astrometric size of Lense-Thirring precession in Kerr and KS geometry against semi-major axis ($a'$) for different mass of scalarons. Orbits with semi-major axes of $45$ au, $100$ au \& $1000$ au (S2 like orbit) are marked by three vertical lines.}
\label{fig1}
\end{figure*}

Here, $\Omega^{KS}_{LT}$ has dimensions of time$^{-1}$. The well known Kerr limit of the LT frequency is natural outcome of infinite scalaron mass in (\ref{eq71}) and is given by,

\begin{equation}\label{eq72}
    \Omega^{K}_{LT}=\frac{mac}{\rho^3(\rho^2-2mr)}\left[4 r^2 \Delta \cos^2\theta+\left(\rho^2-2r^2\right)^2\sin^2\theta\right]^{1/2}
\end{equation}

where, $\Delta$ reduces to $\Delta=r^2-2mr+a^2$, that of Kerr metric. Also, in the weak field limit (i.e. $r>>m$) with $m \propto 1/M_\psi$, equation (\ref{eq70}) reduces to the well known form,

\begin{equation}\label{eq73}
    \vec{\Omega}_{LT}=\frac{GJ}{c^2r^3}\left[2\cos\theta \hat{r}+\sin\theta\hat{\theta}\right]
\end{equation}

which can be deduced from equation (\ref{eq69}). Using equations (\ref{eq71}) and (\ref{eq72}) the LT precession in different orbits have been estimated. To compare the LT precession with the capability of existing and upcoming observing facilities, the astrometric shift of LT precession ($\Omega_{LT}^j\times \frac{a'}{D}\sin(i)$, where $j=K,KS$ and orbital inclination $i$ is chosen as $90^o$) has been estimated in $\mu$as/yr. The LT precession is measured at the perihelion ($r_p=a'(1-e)$) for different orbits. High eccentricity ($e=0.9$) of the orbits having semi-major axes having $a'=(45-1000)$ au is considered.The entire range of scalaron mass ($10^{-22}$ eV - $10^{-16}$ eV) is taken into account. The variation of astrometric size of LT precession against orbital size is presented in Figure  \ref{fig1}.

\begin{figure}
    \centering
    \includegraphics[width=\columnwidth]{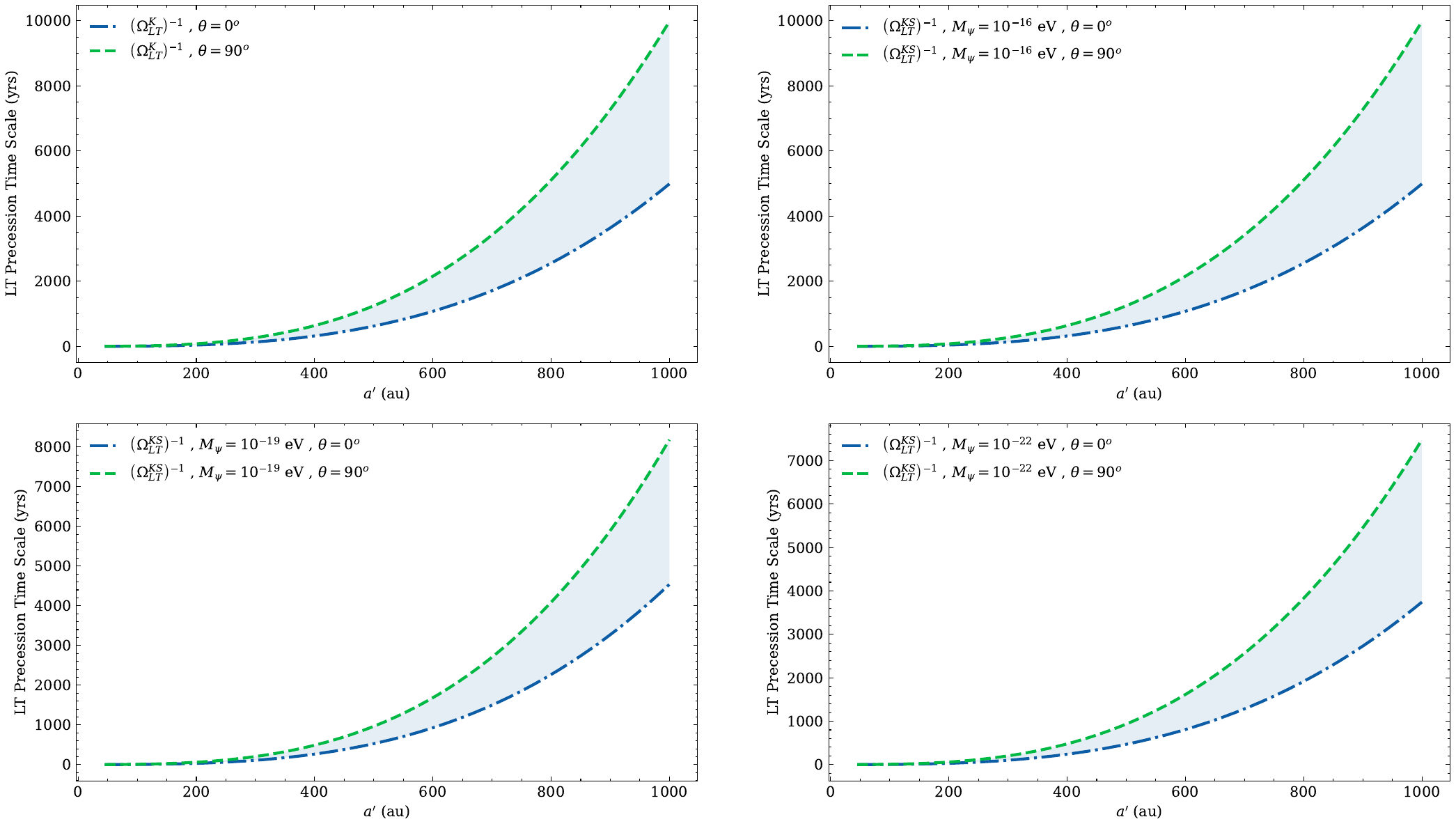}
    \caption{Variation of LT precession time scale in Kerr and KS geometry ($10^{-16}$ eV, $10^{-19}$ eV and $10^{-22}$ eV) against semi-major axis ($a'$) plotted for $e=0.9$.}
    \label{fig5}
\end{figure}

From Figure \ref{fig1}, it has been observed that for the heavier scalarons ($10^{-17}$ eV \& $10^{-16}$ eV) the LT precession in Kerr and KS geometry become degenerate. In the orbit of S2, $\Omega_{LT}^K$ is estimated to be $24.49\ \mu$as/yr for $\theta=0^o$. On the other hand, $\Omega_{LT}^{KS}$ falls in the range ($32.65-24.49$) $\mu$as/yr for the mass of scalarons in the range ($10^{-22}-10^{-16}$) eV respectively. For $\theta=90^o$, $\Omega_{LT}^K$ is found to be $12.25\ \mu$as/yr and $\Omega_{LT}^{KS}$ is found to be in the range ($16.34-12.25$) $\mu$as/yr. This result shows that with one full period of S2 like orbits the LT precession will be capable of constraining scalaron mass through astrometric capability of GRAVITY interferometer in VLT($\sim 30 \ \mu$as). Also, \citet{Fragione_2022} reported that the high spin of the GC black hole could result in the LT precession with time scale shorter than the average age of S-stars ($\sim 10$ Myr, see \citet{Habibi_2017}). The LT precession time scale ($(\Omega_{LT}^j)^{-1}$ in yrs; $j=K,KS$) for the same range of semi-major axis ($45-1000$) au and eccentricity, $e=0.9$ is estimated. The variation of LT precession time scale with respect to semi-major axes for three scalaron masses ($10^{-22},10^{-19}\ \& 10^{-16}$) eV and ($\theta=0^o-90^o$) is displayed with GR (Kerr metric) time scales in Figure \ref{fig5}. It is seen that these time scales for both Kerr and KS metric are far below the average age of S-stars ($\sim 10$ Myr). Therefore, this result may call for higher spin value of GC black hole.

\subsection{Analysis of the orbit equation}\label{sec3.3}

The additional Yukawa like terms ($e^{-M_\psi r}/r$) in the orbit equation (\ref{eq52}) are expected to affect the orbit of a test particle. Now, the orbit equation under the assumption $E<<c^2$ reduces to

\begin{multline}\label{eq53}
u''+u=\underbrace{\frac{mc^2}{L^2}}_\textrm{\tiny{I}}+\underbrace{3mu^2}_\textrm{\tiny{II}} -\underbrace{\frac{8m^2ac^3}{L^3}u}_\textrm{\tiny{III}} +\underbrace{\frac{mc^2}{3L^2}e^{-\frac{M_\psi}{u}}}_\textrm{\tiny{IV}}+\underbrace{mu^2e^{-\frac{M_\psi}{u}}}_\textrm{\tiny{V}}\\
-\underbrace{\frac{8m^2ac^3}{L^3}\left(\frac{2}{3}ue^{-\frac{M_\psi}{u}}+\frac{1}{9}ue^{-\frac{2M_\psi}{u}}+\frac{1}{9}M_\psi e^{-\frac{2M_\psi}{u}}+\frac{1}{3}M_\psi e^{-\frac{M_\psi}{u}}\right)}_\textrm{\tiny{VI}}+\underbrace{\frac{mc^2M_\psi}{3L^2}u^{-1}e^{-\frac{M_\psi}{u}}}_\textrm{\tiny{VII}}+\underbrace{\frac{m M_\psi}{3}u e^{-\frac{M_\psi}{u}}}_\textrm{\tiny{VIII}}
\end{multline}

From equation (\ref{eq53}) it is seen that there are several correction terms whose effect in the orbit is unknown. Even though the mass of the scalarons is constrained through shadow measurement (section \ref{sec3.1}) and pericentre shift (section \ref{sec3.2}), we wish to explore the nature of these correction terms in the entire spectrum of scalaron mass ($10^{-22}$ eV - $10^{-16}$ eV). The individual terms in the orbit equation are displayed below.

\begin{empheq}[right=\empheqrbrace]{equation}\label{eq54a}
\begin{aligned}
& \text{I}= \text{Newtonian part}, \text{II}=\text{Schwarzschild part},\\
& \text{III}=\text{Kerr part}, \text{IV}=\text{Scalaron modification to Newtonian part},\\
& \text{V}=\text{Scalaron modification to Schwarzschild part},\\
& \text{VI}=\text{Scalaron modification to Kerr part},\\
& \text{VII \& VIII}=\text{Scale modulated scalaron correction}
\end{aligned}
\end{empheq}

 The effect of these corrections in the orbit of any test particle has been analysed by estimating the ratio of each of the correction terms to their GR and Newtonian counterparts. These ratios are taken as,

\begin{empheq}[right=\empheqrbrace]{equation}\label{eq54}
\begin{aligned}
& \frac{\text{IV}}{\text{I}}=\frac{\text{Scalaron modification to Newtonian part}}{\text{Newtonian part}}\\
& \frac{\text{V}}{\text{II}}=\frac{\text{Scalaron modification to Schwarzschild part}}{\text{Schwarzschild part}}\\
& \frac{\text{VI}}{\text{III}}=\frac{\text{Scalaron modification to Kerr part}}{\text{Kerr part}}
\end{aligned}
\end{empheq}

\begin{figure*}
\begin{center}
    \begin{minipage}[b]{0.7\columnwidth}
        \includegraphics[width=\columnwidth]{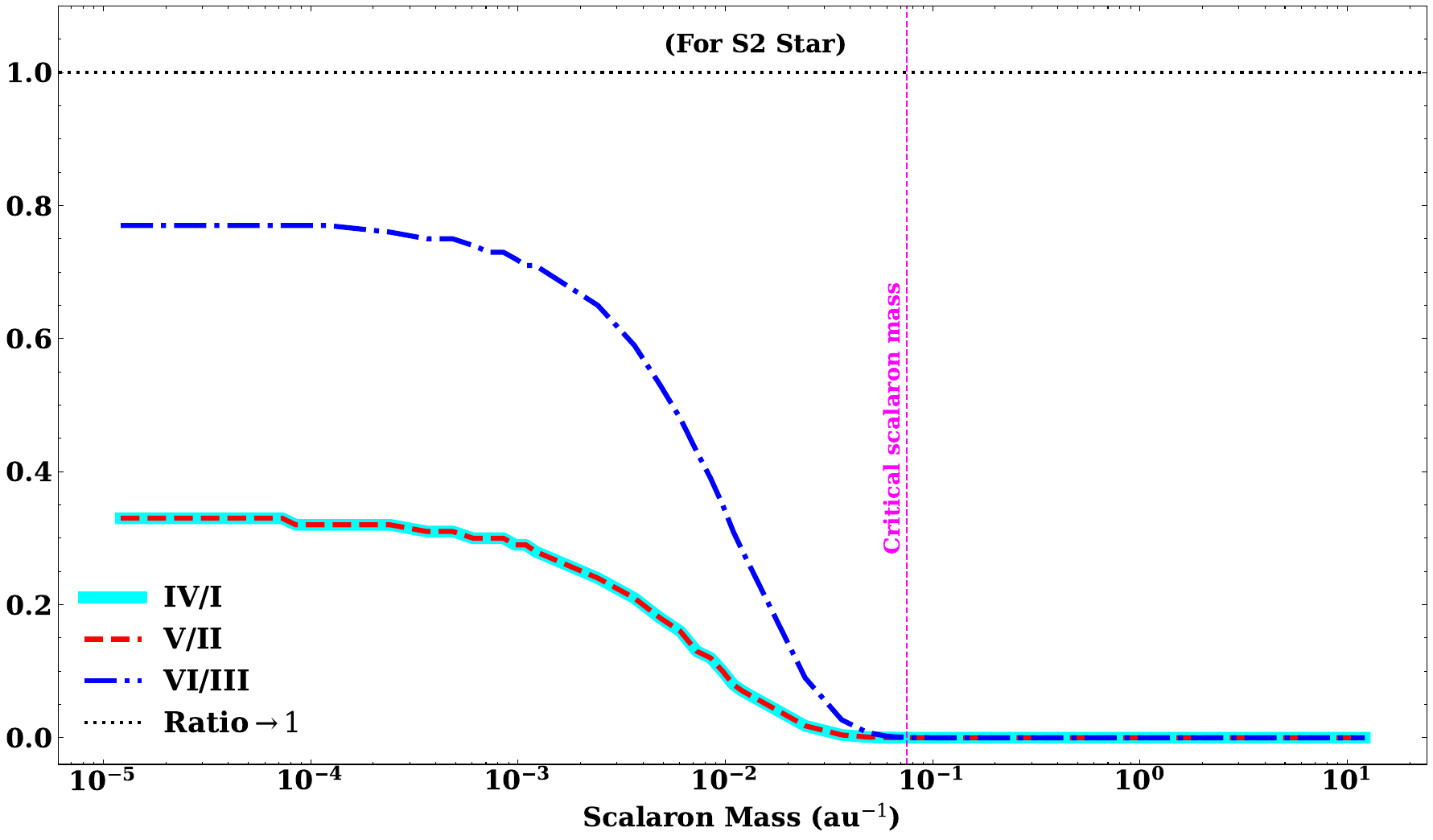}
    \end{minipage}\\ \vspace{0.5cm}
    \begin{minipage}[b]{0.7\columnwidth}
        \includegraphics[width=\columnwidth]{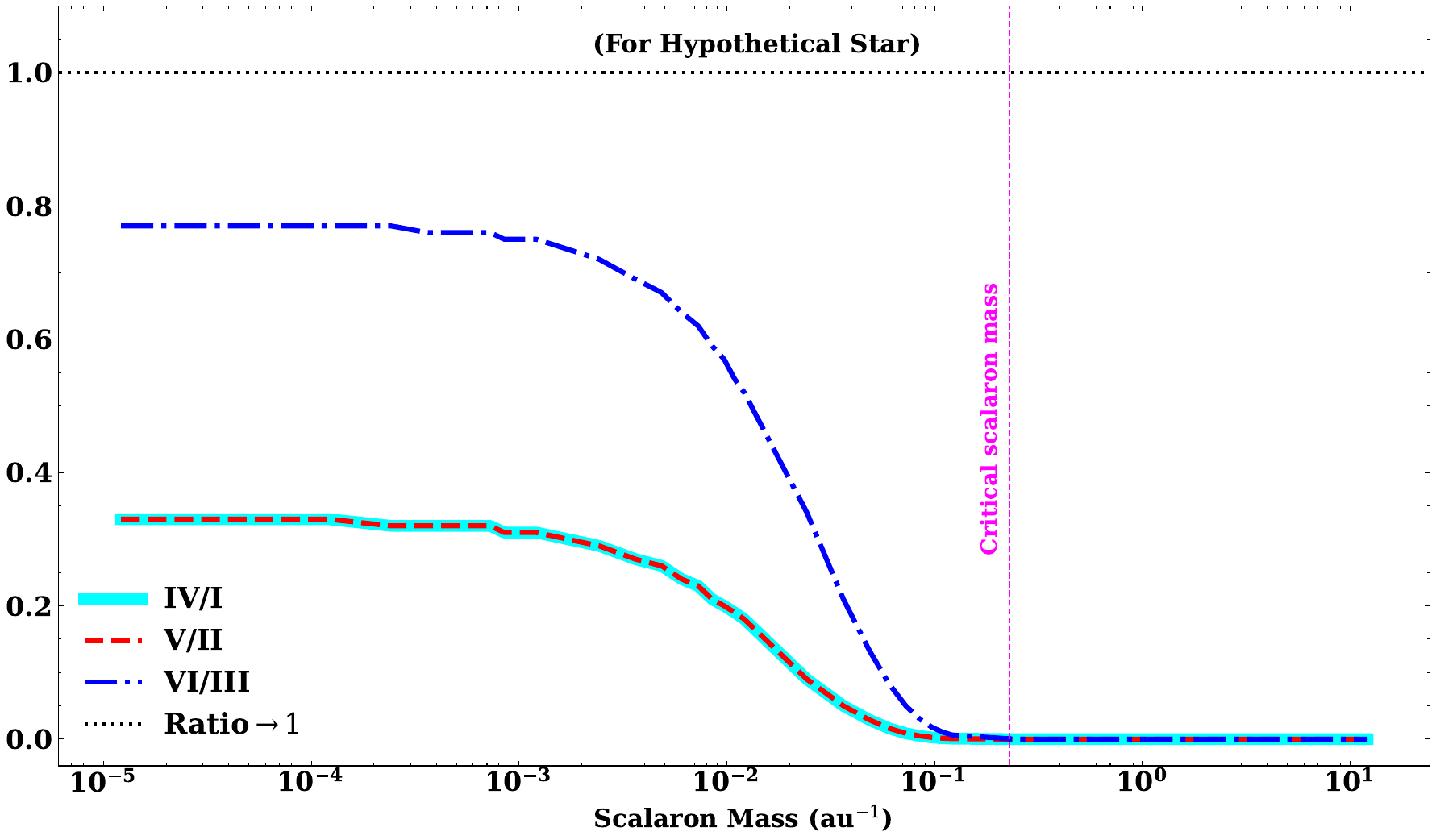}      
    \end{minipage}
\end{center}
\caption{The variation of the ratios of different correction terms to their GR and Newtonian counterparts appearing in the orbit equation against mass of the scalarons in the orbit of S2 and the hypothetical star ($r=50$ au (pericenter distance), $a'=500$ au and $e=0.9$). The horizontal dotted line shown in both the panels corresponds to unit value of the ratio of modified gravity effects to their GR and Newtonian counterparts}.
\label{fig2}
\end{figure*}

The above three ratios are estimated for the orbit of the star S2 (having $r=120$ au (pericenter distance) and semi-major axis, $a'=970$ au) as well as for a hypothetical star (with $r=50$ au (pericenter distance), $a'=500$ au and $e=0.9$). These variations are presented in Figure \ref{fig2}. It is observed that in the orbit of S2, the scalaron effect becomes negligible around $0.074$ au$^{-1}$($\approx 10^{-19}$ eV). Similarly, in the orbit of hypothetical star, this effect becomes negligible around $0.22$ au$^{-1}$($\approx 10^{-18}$ eV). 

\begin{figure*}
\begin{center}
    \begin{minipage}[b]{0.48\columnwidth}
        \includegraphics[width=\columnwidth]{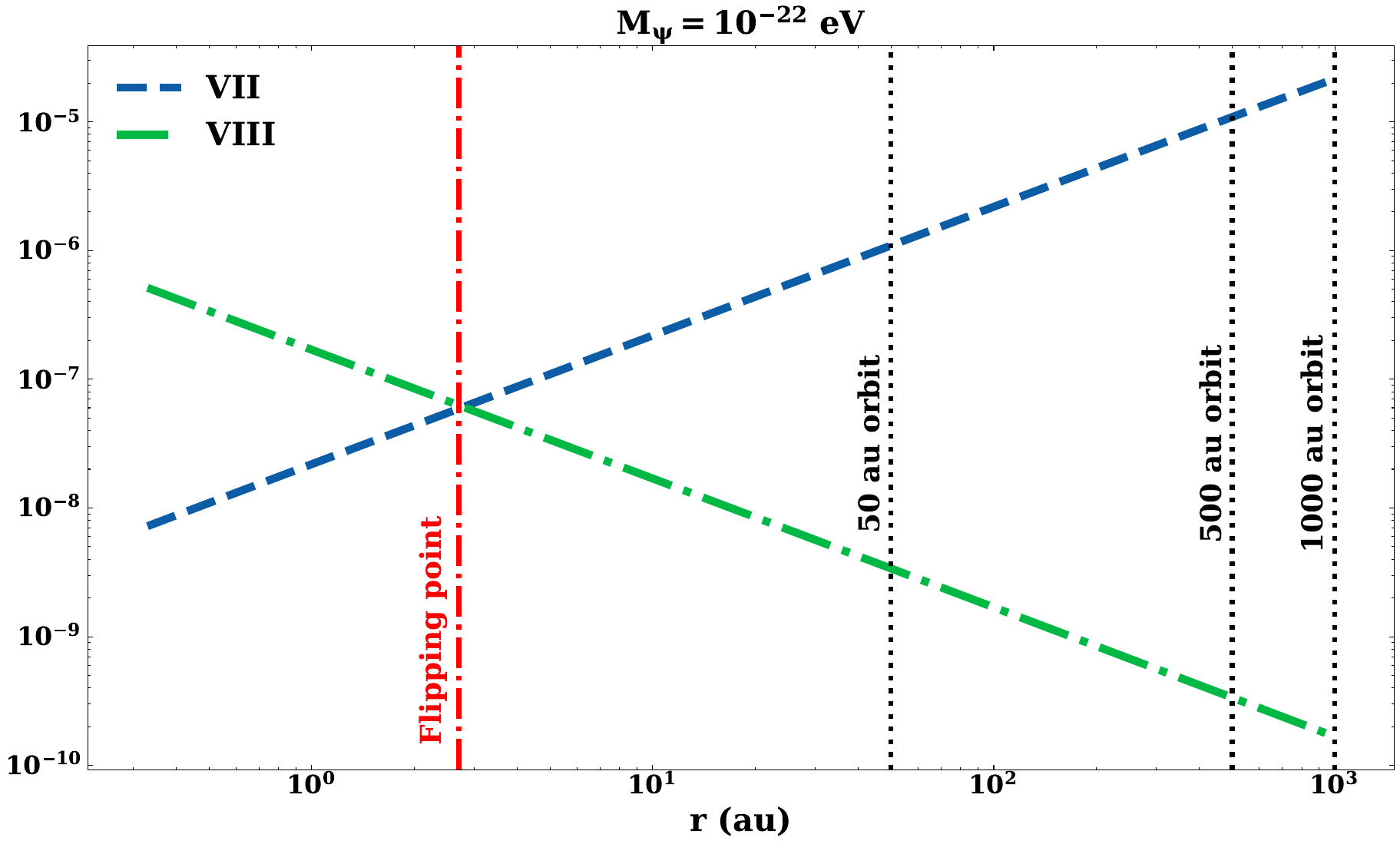}
    \end{minipage}
    \begin{minipage}[b]{0.48\columnwidth}
        \includegraphics[width=\columnwidth]{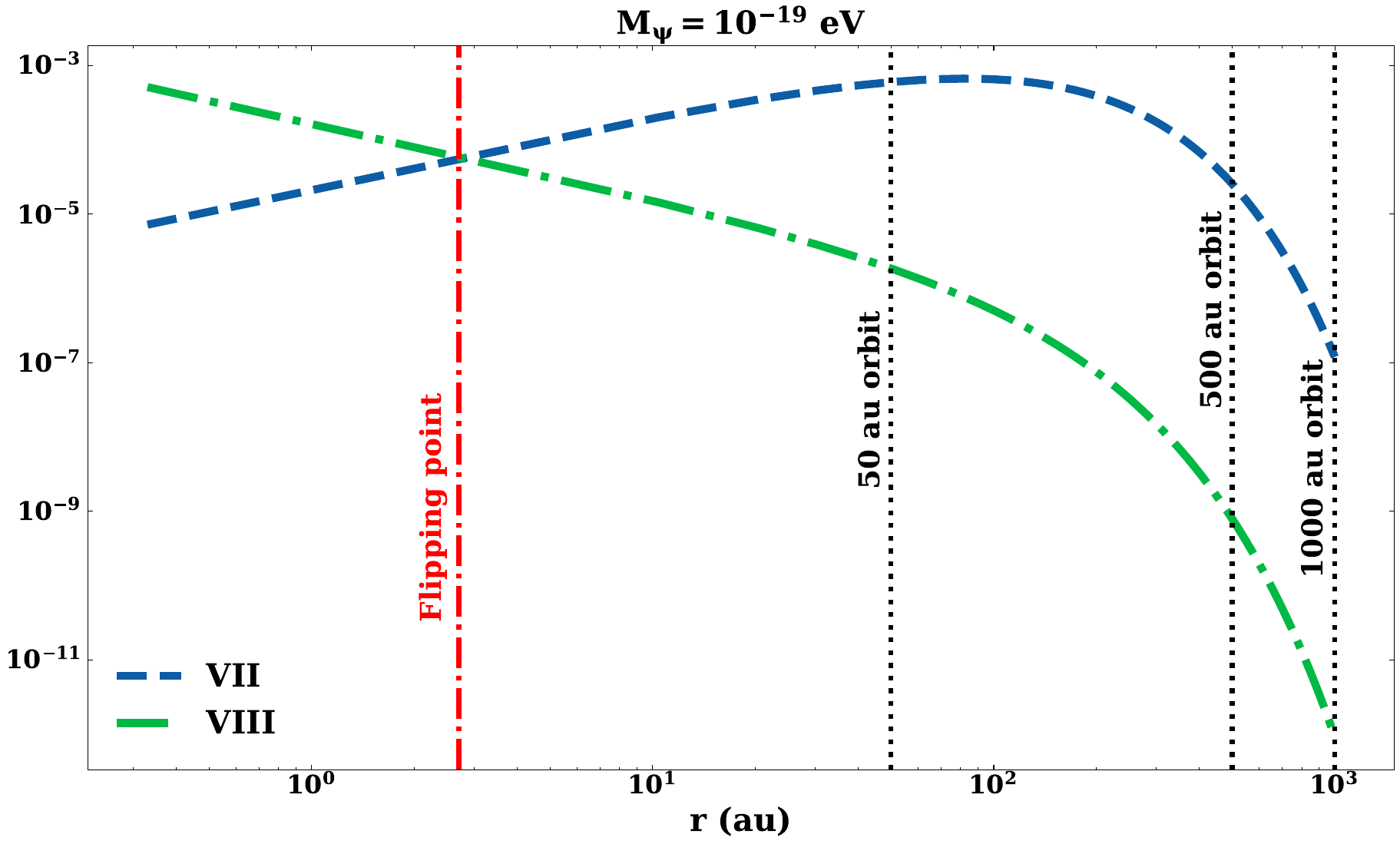}
    \end{minipage}\\ \vspace{0.3cm}
    \begin{minipage}[b]{0.48\columnwidth}
        \includegraphics[width=\columnwidth]{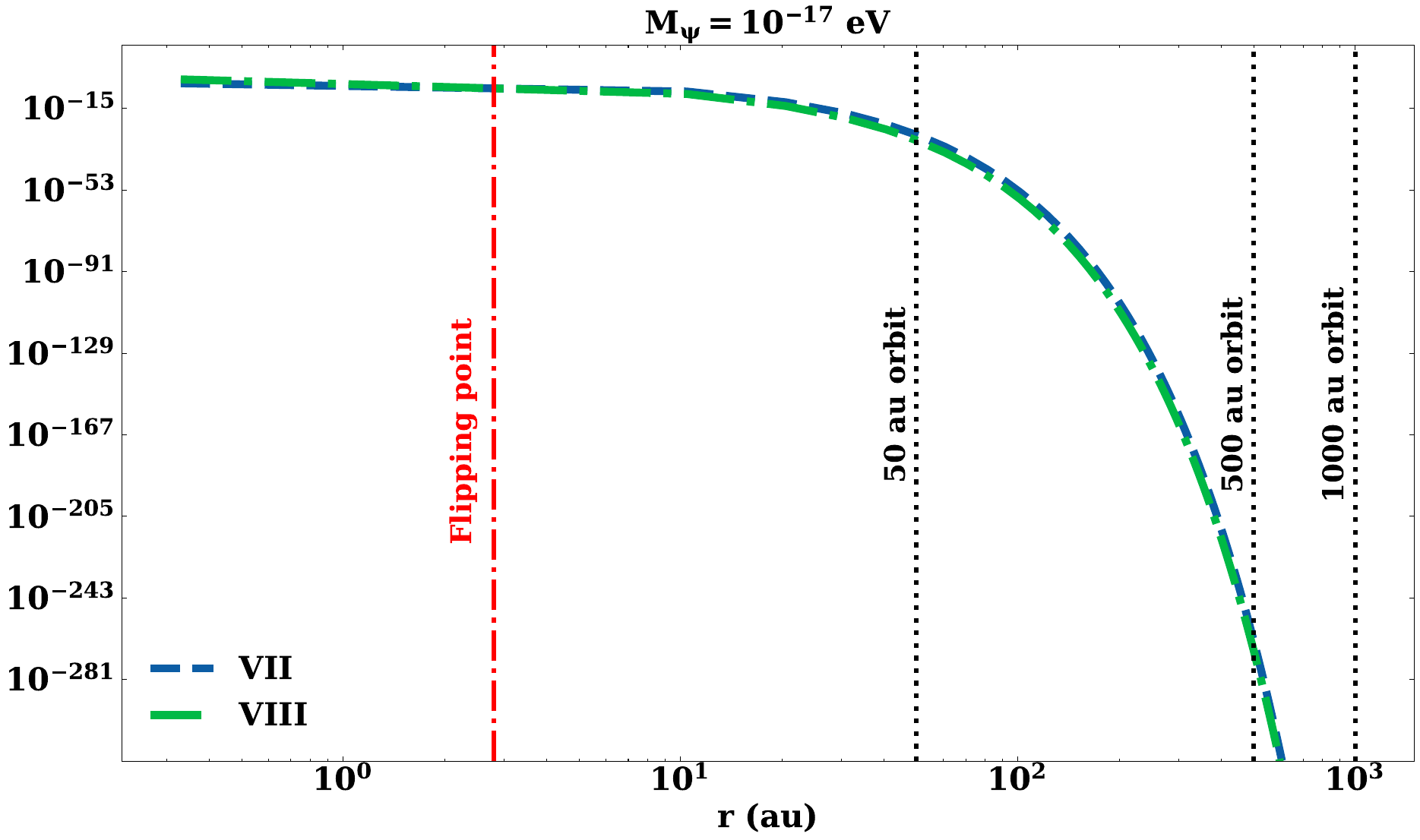}
    \end{minipage}
\end{center}
\caption{The variation of scale modulated terms (term VII and VIII) in orbit equation (\ref{eq53}) against orbital scale ($r$) for different scalaron masses. Orbits with $r=50$ au, $500$ au and $1000$ au are marked with three vertical lines}.
\label{fig3}
\end{figure*}

The scale modulated scalaron correction terms are,

\begin{equation}\label{eq55}
\begin{aligned}
& \text{VII}=k r e^{-M_\psi r};
& \text{VIII}=k' r^{-1} e^{-M_\psi r}
\end{aligned}
\end{equation}

where, $k=mc^2M_\psi/3L^2$ and $k'=mM_\psi/3$. The effect of these two terms in the orbit of a test star is analysed separately. The size of the EHT shadow $48.7\pm 7.0\ \mu$as \citep{Akiyama_VI} corresponds to a range of $0.33$ au to $0.45$ au. Thus, variation of these two terms has been investigated for the range $r=0.33$ au - $r=1000$ au. The variation of these terms in this range is investigated for three scalaron masses: $10^{-22}$ eV, $10^{-19}$ eV and $10^{-17}$ eV so as to cover the entire scalaron mass spectrum\footnote{ The variation for scalaron mass $10^{-16}$ eV is not discussed, as for these scalarons the two terms become exactly same.}. The variations are presented in Figure \ref{fig3}. For $10^{-22}$ eV and $10^{-19}$ eV scalarons it is seen that the term VII dominates over the term VIII for orbits with $r=2.7$ au and higher. But the terms change role at $r=2.7$ au and below (we refer to it as "Flipping point"). For $10^{-17}$ eV scalarons it is difficult to distinguish between these two terms and they are of extremely tiny magnitude at all orbital radii which is consistent with massive scalarons showing GR like effects.

\section{Results and discussions}\label{results}

In this work, we constructed an axially symmetric and asymptotically flat metric called as the Kerr-scalaron (KS) metric by employing Newman-Janis algorithm (NJA) to the spherically symmetric Schwarzschild-scalaron (SchS) metric of f(R) gravity. The KS metric is used to construct effective potential and orbit equation and to estimate pericenter shift and black hole shadow size. The KS metric reduces to Kerr metric for infinitely large scalaron mass and it reduces to Schwarzschild metric for infinitely large scalaron mass and vanishing spin parameter. The asymptotic flatness of the KS metric is found to be a natural outcome of an inverse relationship between scalaron mass ($M_\psi$) and black hole mass ($m$). The effective potential and orbit equation have appropriate Kerr and Schwarzschild limits. The absence of orbital degeneracy ($\omega_r \neq \omega_\phi$) in the metric ensures existence of scalaron induced pericenter shift which reduces to standard general relativistic values of pericenter shift under the limits $M_\psi \rightarrow \infty$ and ($M_\psi \rightarrow \infty,a\rightarrow 0$).

\par The angular size of the GC black hole shadow in KS geometry is investigated for the scalaron mass range ($10^{-22}$ eV - $10^{-16}$ eV). The observed angular size of the bright emission ring of the GC black hole shadow has been utilized to estimate deviation of the KS metric from general relativity. It is found that in the Schwarzschild limit of the KS metric, the deviation parameter for all scalaron masses is outside the bound reported by EHT observations. However, for the KS metric with $\chi=0.90$ and $\theta=0^o$ the deviation parameter for massive scalarons ($10^{-17}$ eV \& $10^{-16}$ eV), is within the EHT bounds. These are obtained as $\delta\approx -0.009$ and $\delta\approx-0.019$ for scalaron mass $10^{-17}$ eV and $10^{-16}$ eV respectively.

\par The effect of scalarons on the pericenter shift of stellar orbits near the GC black hole is investigated. For this study the deviation of Schwarzschild limit of KS pericenter shift $(\delta\phi)_{SchS}$ from standard Schwarzschild pericenter shift $(\delta\phi)_{Sch}$ is calculated for the scalaron masses $10^{-17}$ eV and $10^{-16}$ eV. From the variation of this deviation plotted with respect to semi-major axis in the range ($a'=45.40$ au - $1000$ au) it is found that scalarons with mass $10^{-16}$ eV almost produce standard Schwarzschild pericenter shift at all orbits. But $10^{-17}$ eV scalarons dominate standard Schwarzschild precession for very compact orbits ($a'<<50$ au). The deviation reaches a magnitude of around $11$ arc minute for orbits just below $45$ au. For the orbits of S2 and S4716 both these scalarons show standard Schwarzschild precession (see figure \ref{fig4}). The $f_{SP}$ parameter which measures the deviation from Schwarzschild precession is also investigated. The deviation $(\delta\phi)_{SchS}-(\delta\phi)_{Sch}$ has been estimated from Figure \ref{fig4} for three hypothetical orbits, $a'=45$ au, $50$ au, $100$ au. This has helped us to estimate $f_{SP}$ values which are displayed in table \ref{tab1}. The Schwarzschild precession for the three orbits has been estimated for S2 like eccentricity ($e=0.9$). The scalaron mass required for the calculation is chosen as $10^{-17}$ eV. From table \ref{tab1} it is found that the $f_{SP}$ values are quite close to general relativistic value. Therefore, we do not notice drastic deviation of f(R)-scalaron gravity from GR down to orbital scale of $45$ au. Even though stellar orbits below $45$ au are not yet understood, we infer that massive scalarons have the capability to affect pericenter shift far below $45$ au (see figure \ref{fig4}). The deviation in the pericenter shift ($11$ arc minute and more) is measurable through existing astrometric facilities available in the Very Large Telescope (VLT) which has measured pericenter shift of S2 ($\sim 12$ arc minute).

\par The Lense-Thirring effect in KS geometry is explored for the scalaron mass range ($10^{-22}$ eV - $10^{-16}$ eV). For the orbit of S2, the astrometric size of LT precession is obtained in the range ($32.65-24.5$) $\mu$as/yr for $\theta=0^o$ and ($16.34-12.25$) $\mu$as/yr for $\theta=90^o$. These ranges of LT precession are accessible through astrometric capabilities of GRAVITY interferometer in VLT. Scalarons with $10^{-17}$ eV and $10^{-16}$ eV produce LT precession of same size as that of GR for all orbital scales considered. The ultra light scalarons ($10^{-22}$ eV), however, produce LT frequency which is larger than that of GR by around $33\%$ in all the orbits. Therefore, LT precession constrains the scalaron mass which is compatible with the mass constrained by black hole shadow and Schwarzschild precession of S2. The maximum value of LT precession time scale for scalarons with $M_\psi=10^{-22}$ eV and $M_\psi=10^{-19}$ eV is only $20\%-25\%$ away from that of GR and $10^{-16}$ eV scalarons ($10000$ yrs). This time scale is only about $0.1 \ \%$ of the age of the S-stars. Scalarons are found to reduce LT precession time scale. This indicates that the GC black hole preferably has higher value of spin which complements the result obtained through the deviation parameter extracted from ring diameter of the black hole shadow.

\par The ratios of scalaron induced correction terms to their GR and Newtonian counterparts are analysed. The analysis has been carried out in the orbit of the star S2 ($r=120$ au, $a'=970$ au) and a hypothetical star ($r=50$ au, $a'=500$ au) for scalaron mass ranging from $10^{-22}$ eV to $10^{-16}$ eV.
In the orbit of S2, it has been found that the effect of scalarons is significant only below $10^{-19}$ eV. In case of the hypothetical star with compact orbit the scalaron effect is found to be prominent for mass below $10^{-18}$ eV. In both these cases, for scalarons having mass above $10^{-18}$ eV the effect of modified gravity is heavily suppressed by GR. From the analysis of the orbit equation (\ref{eq53}) it is found that for low mass scalarons ($10^{-22}$ - $10^{-19}$) eV there exists a scale modulated scalaron gravity correction $kr e^{-M_\psi r}$ (term VII) which shows effect on wider orbits. But for high mass scalaron with $M_\psi=10^{-17}$ eV effect of such correction is heavily suppressed by GR.

\section{Conclusion}\label{summary}
The Kerr-scalaron metric of f(R)-gravity has been found to possess appropriate GR limit. The asymptotic flatness of the metric has been ensured by an inverse relationship between scalaron mass and black hole mass. Since scalaron mass is not independent of the black hole mass, presence of scalarons near black hole restores the no-hair theorem of general relativity. From the astronomical consequences of the metric it has been found that massive scalarons ($10^{-17}$ eV and $10^{-16}$ eV) produce GR like observables, for example pericenter shift down to 45 au, LT precession frequency near S2 like orbits and black hole shadow. The astrometric capabilities of existing large telescopes and upcoming ELTs are expected to be capable enough for testing the new metric and its astronomical consequences within the regime of very compact orbits near the GC black hole.

%\section{Floats} \label{sec:floats}

%% Also note that the akcnowlodgment environment does not support long amounts of text. If you have a lot of people and institutions to acknowledge, do not use this command. Instead, create a new \section{Acknowledgments}.
\begin{acknowledgments}
The corresponding author acknowledges Sajanth Subramaniam, ETH Z\"{u}rich for his helpful suggestions on the orbit equation during a private communication.
\end{acknowledgments}

\bibliography{references}{}
\bibliographystyle{aasjournal}



\end{document}